%% file: moving-bmu.tex
\documentclass[a4paper,10pt,twocolumn]{article}

\usepackage{amsmath}
\usepackage{graphicx}
\usepackage{xcolor}
\usepackage{url}
\usepackage[colorlinks=true,linkcolor=blue,citecolor=blue]{hyperref}
\usepackage[charter]{mathdesign}
\usepackage[expproduct=cdot]{siunitx} 
\usepackage[font=footnotesize,labelfont=bf,labelsep=endash,margin=0pt]{caption} 

\newcommand{\ie}{{\it i.e.}}
\newcommand{\eg}{{\it e.g.}}
\newcommand{\etal}{{\it et\,al.}}
\newcommand{\etc}{{\it etc}}

\newcommand{\vs}{\textit{vs.}}

\renewcommand{\d}{\mathrm{d}}
\newcommand{\p}{\partial}
\newcommand{\pd}[2]{\frac{\partial #1}{\partial #2}}

\newcommand{\Order}{\mathrm{O}}

\newcommand{\e}{\mathrm{e}}
\newcommand{\avg}[1]{\langle #1 \rangle}

\renewcommand{\b}[1]{{\boldsymbol{#1}}} 

\newcommand{\bmu}{\textsc{bmu}} 

\newcommand{\ob}{\textsc{ob}}

\newcommand{\obu}{\text{\textsc{ob}$_\text{u}$}} 

\newcommand{\obp}{\text{\textsc{ob}$_\text{p}$}}
\newcommand{\oba}{\text{\textsc{ob}$_\text{a}$}}

\newcommand{\ocp}{\text{\textsc{oc}$_\text{p}$}}

\newcommand{\oca}{\text{\textsc{oc}$_\text{a}$}}
\newcommand{\ocm}{\text{\textsc{oc}$_\text{m}$}}
\newcommand{\tgfb}{\textsc{tgf-$\beta$}}
\newcommand{\Tgfb}{\textsc{Tgf-$\beta$}}

\def\dkk1{\text{\textsc{d}kk$_1$}}

\newcommand{\rank}{\textsc{rank}}
\newcommand{\rankl}{\textsc{rankl}}
\newcommand{\Rankl}{\textsc{Rankl}}
\newcommand{\opg}{\textsc{opg}}

\def\pgde2{\text{\textsc{pgde}$_2$}}
\newcommand{\pth}{\textsc{pth}}

\def\lrp5{\text{\textsc{lrp}$_5$}}
\newcommand{\piact}{\ensuremath{\pi^\text{act}}} 
\newcommand{\pirep}{\ensuremath{\pi^\text{rep}}} 

\newcommand{\kres}{\ensuremath{k_\text{res}}}

\newcommand{\da}{\ensuremath{\text{day}}} 

\newcommand{\um}{\ensuremath{\muup\text{m}}} 

\begin{document}

    \title{{\bf\vskip-9mm Spatio-temporal structure of cell distribution in cortical Bone Multicellular Units: a mathematical model}}

    \author{P R Buenzli, P Pivonka and D W Smith}
    \date{\small \vspace{-2mm}Engineering Computational Biology Group, FECM, The University of Western Australia, WA 6009, Australia\\\vskip 1mm \normalsize \today\vspace*{-5mm}}
    \twocolumn[
    \vskip-5mm
    \begin{@twocolumnfalse}
        \maketitle
        \begin{abstract}
Bone remodelling maintains the functionality of skeletal tissue by locally coordinating bone-resorbing cells (osteoclasts) and bone-forming cells (osteoblasts) in the form of Bone Multicellular Units (\bmu s). Understanding the emergence of such structured units out of the complex network of biochemical interactions between bone cells is essential to extend our fundamental knowledge of normal bone physiology and its disorders. To this end, we propose a spatio-temporal continuum model that integrates some of the most important interaction pathways currently known to exist between cells of the osteoblastic and osteoclastic lineage. This mathematical model allows us to test the significance and completeness of these pathways based on their ability to reproduce the spatio-temporal dynamics of individual \bmu s. We show that under suitable conditions, the experimentally-observed structured cell distribution of cortical \bmu s is retrieved. The proposed model admits travelling-wave-like solutions for the cell densities with tightly organised profiles, corresponding to the progression of a single remodelling \bmu. The shapes of these spatial profiles within the travelling structure can be linked to the intrinsic parameters of the model such as differentiation and apoptosis rates for bone cells. In addition to the cell distribution, the spatial distribution of regulatory factors can also be calculated. This provides new insights on how different regulatory factors exert their action on bone cells leading to cellular spatial and temporal segregation, and functional coordination.

\vspace{2mm}\noindent\textbf{Keywords:} bone cell interactions, cortical BMU, spatio-temporal bone remodelling, \rank--\rankl--\opg, mathematical modelling
        \end{abstract}
        \vspace{5mm}
    \end{@twocolumnfalse}
    ]%

\section{Introduction}
\begin{figure}\begin{center}
        \includegraphics[width=1\columnwidth]{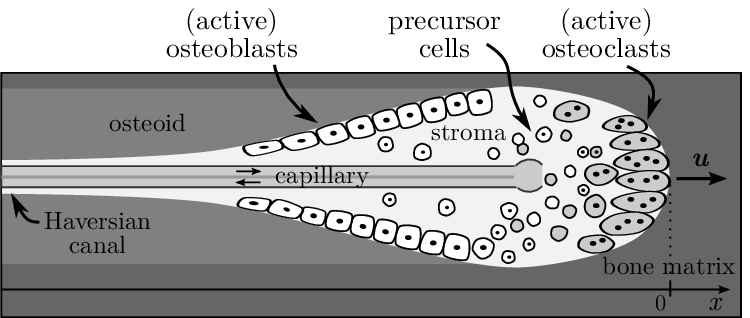} 
        \includegraphics[width=1\columnwidth]{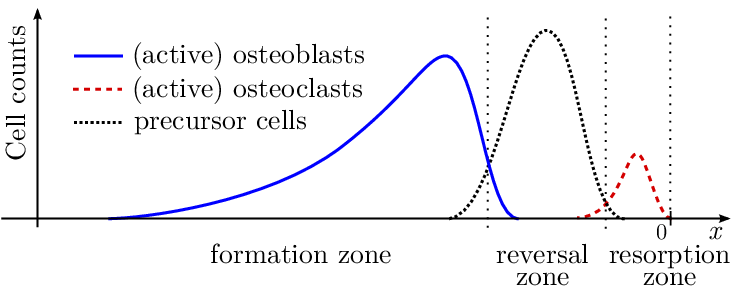}
        \caption{Schematic figure of the internal organisation of a cortical \bmu. Osteoclasts resorb the bone matrix at the front while osteoblasts lay down osteoid towards the back to refill the cavity. The central capillary provides a supply of precursor cells, as well as various nutrients. A schematic plot of the number of cells present at each position $x$ along the \bmu\ is depicted below.}
        \label{fig:schematic-bmu}
        \vspace{-5mm}
\end{center}\end{figure}

In human adults, between 5 and 30\% of bone volume is replaced every year \cite{martin-burr-sharkey,bone-mechanics-handbook} in a process referred to as remodelling. Bone replacement is accomplished by stand-alone groups of cells of the osteoclastic and osteoblastic lineage progressing through old bone over a period of several weeks. Such a group of cells is called a ``Bone Multicellular Unit'' (\bmu) and can be viewed as the basic functional unit for bone remodelling \cite{frost2,frost3,parfitt1,parfitt2}. Tetracycline-based histomorphometry has considerably helped in the elucidation of the spatial organisation and kinetic properties of the different bone cells in cortical \bmu s \cite{parfitt3,jaworski-hooper,jaworski-duck-sekaly}, which clearly indicates a spatial segregation of bone cell types depending on cell maturity. At the front of a \bmu, in a region called the resorption zone (see Figure~\ref{fig:schematic-bmu}), active osteoclasts attach to the bone surface and dissolve bone by secreting a mixture of proteases that break down the collagenous matrix, and hydrogen ions that reduce the pH and dissolve the minerals into the micro-environment \cite{vaananen-etal,roodman}. Towards the back of the \bmu, in the so-called formation zone, active osteoblasts refill the cavity by  laying down a collagen-rich substance known as osteoid, which subsequently mineralises to form new bone over the following month or so (see \cite{parfitt3,martin-burr-sharkey,bone-mechanics-handbook}). The region between the resorption zone and the formation zone, referred to as the reversal zone, contains precursor cells of both lineages embedded in a loose connective tissue stroma \cite{parfitt3}. New precursor cells and nutrients are supplied to the \bmu\ by a small capillary that grows at the same rate as the \bmu\ progresses into the bone. The net effect of the passage of a \bmu\ at a specific location of bone is the local renewal of the bone matrix and the formation of a so-called ``secondary osteon'', which includes a new Haversian canal.

The existence of such a functional remodelling unit (referred to by Frost as a ``packet of turnover'' \cite{frost2}) suggests the presence of tight couplings between the various cell types composing \bmu s. It has been hypothesised several decades ago that some combination of local and/or systemic signals structure this internal cellular organisation \cite{frost2,parfitt2}. In the mid 1990s, the discovery of the \rank-\rankl-\opg\ pathway explained many previous experimental observations. This regulatory pathway can be expected to play a major role in \bmu\ physiology. Many other potential regulatory molecules have been found by experimental biologists (including systemic hormones, nerve signals, vascular agents, growth factors, chemokines, \etc; see \cite{martin,lemaire-etal,pivonka-etal1}). However, it is yet to be proven that these local interactions are able to group several generations of osteoclasts and osteoblasts in the form of \bmu s that present a clear spatial and temporal separation of these cellular activities. While the structure of \bmu s is well understood at a descriptive level \cite{parfitt3,martin-burr-sharkey,bone-mechanics-handbook}, how this structure is linked to the fundamental underlying cellular interaction mechanisms remains to be elucidated. The present work aims to address this question.

In this paper, we extend our previous temporal model of bone remodelling \cite{pivonka-etal1} into a one-dimensional spatio-temporal model. Using this model, we study how bone cells structure themselves into a cortical \bmu\ under the action of intercellular signalling. This model is based on fundamental material-balance equations expressed as partial differential equations (PDEs). Non-conservative production or elimination of biochemical components in these general continuity equations are prescribed in accordance with the known biochemistry currently believed to play the most important role in bone remodelling. As such, the model explicitly includes transforming growth factor $\beta$ (\tgfb), parathyroid hormone (\pth) and the receptor--activator nuclear factor $\kappa\beta$ axis consisting of the receptor \rank, the ligand \rankl\ and the soluble decoy receptor osteoprotegerin (\opg). These regulatory factors couple two cell types of the osteoclastic lineage (a third one is introduced in Section~\ref{sec:model-structure}) and three cell types of the osteoblastic lineage. Other components of the cellular communication system, known and unknown, are introduced implicitly through various model parameters and external model conditions. For example, the capillary assisting the progression of a cortical \bmu\ is modelled as a localised supply of bone precursor cells around the capillary's (growing) tip. Under these assumptions, we find that the model admits solutions for the cell distributions in the form of travelling waves that have profiles that match the observed internal spatial organisation of a cortical \bmu.

In recent years, several teams of researchers have elaborated mathematical and computational models of bone remodelling, generally monitoring the evolution of the bone cells over time via ordinary differential equations (ODEs) \cite{komarova-etal1,lemaire-etal,pivonka-etal1}. Recently, Ryser \etal\ have included a spatial dimension in the model \cite{komarova-etal1}, addressing the important question of interaction between locally-expressed \rankl\ and soluble \opg\ for a trabecular \bmu\ \cite{ryser-etal1,ryser-etal2}. In their model, \bmu s are driven by a \rankl\ field in the surrounding bone matrix. Other researchers have developed cellular automata simulations to model resorption and formation on a per site basis \cite{huiskes-etal1}.

To our knowledge, no group has yet addressed the issue of internal structuring of cortical \bmu s. Our approach emphasises the detailed integration of the biochemical processes involving osteoclastic and osteoblastic cells at several maturation stages into a comprehensive partial differential model of the cortical \bmu. Since it is based on a general formulation of the material-balance equation, the construction of the model is modular and extensible. New interaction pathways or cell types can be included as needed. The one-dimensional continuous formulation employed here enables us to investigate analytically how the various cell distributions making the internal structure of the \bmu\ depend on the model assumptions.

The paper is organised as follows: the model formulation is described in Section~\ref{sec:model}. In Section~\ref{sec:density-profiles}, the system of coupled nonlinear PDEs is then solved numerically for the various cell and regulatory factor distribution profiles along the \bmu. Theoretical investigations of these profiles are performed, allowing us to map some of the profiles' properties to parameters of the model. In Section~\ref{sec:model-structure}, we investigate the effects of various model assumptions made in Section~\ref{sec:model}. Finally, we extend the initial model to include a new differentiation stage for osteoclasts, which is required to explain their observed spatial migration from the reversal zone to the resorption zone (see Figure \ref{fig:schematic-bmu}). Concluding remarks are made in Section~\ref{sec:conclusions}.

\section{Mathematical model of cortical {\bmu} remodelling}\label{sec:model}
In the confined environment of a cortical \bmu, the most important phenomena taking place are the biochemical interactions between the cells and their regulatory factors, as well as the directed or diffusive transport of these entities. These phenomena are described in general by the material-balance equations of the species considered~\cite{stewart-lightfoot,evans-morriss,scheurer-stueckelberg}:
\begin{align}\label{mass-balance}
        \pd{}{t} n_A(\b r,t)  =  \sigma_A(\b r, t) - \b \nabla\!\cdot\! \b J_A(\b r, t).
\end{align}
In Eq.~\eqref{mass-balance}, $A$ denotes any cell type or regulatory agent (such as hormones, growth factors, paracrine factors, \etc.) explicitly accounted for in the model; $n_A(\b r, t)$ is the local density or concentration of $A$ (\ie, number of entities $A$ per unit volume)\footnote{To align with common practice, we shall use the terminology ``density'' for cells and ``concentration'' for regulatory factors even if the units are chosen the same.} at point $\b r$ in space and at time t ($\b r$ is a position vector); $\sigma_A(\b r, t)$ denotes local source/sink terms that account for non-conservative mechanisms, such as cellular proliferation, differentiation, apoptosis, or mass action kinetic rates of the regulatory factor binding reactions; $\b J_A(\b r,t)$ is the flux associated with transport properties of $A$, such as diffusion, advection, or resulting from inherent motility, \eg\ chemotaxis. Due to the interactions between cells and regulatory factors, the material-balance equations \eqref{mass-balance} written for all $A$s are coupled. These couplings may arise both through the source/sink terms (\eg\ hormonal up-regulation/down-regulation of a cellular response) and through the fluxes (\eg\ chemotaxis). Note that since the fluxes are differential in space, they are expected to play an important role in the spatial organisation of the cells within the \bmu.

In practice, the definition of local densities relies on a representative volume element large enough to contain many entities, yet small enough to remain local. While only few cells are present in a single \bmu, continuous cellular densities can be defined in a statistical sense \cite{evans-morriss}, \ie, by averaging histograms of cell counts over an ensemble of similar \bmu s (see Figure~\ref{fig:schematic-bmu}).

\subsubsection*{Osteoclasts}
Following the ODE model of bone remodelling proposed by Pivonka \etal\ \cite{pivonka-etal1}, we consider two stages of osteoclast development: ``precursor osteoclasts'' (\ocp s) and ``active osteoclasts'' (\oca s). Precursor osteoclasts are assumed to have derived from hematopoietic progenitor cells and to be delivered to the \bmu\ cavity at the tip of the capillary (see Figure~\ref{fig:schematic-bmu}). In cortical \bmu s, it takes 3 to 4 days for (single-nucleated) pre-osteoclasts to differentiate, migrate and join the dozen or so active multinucleated osteoclasts (each composed of around 10 nuclei) found at the front of the \bmu. These individual nuclei in active osteoclasts are then degraded after around 12 days \cite{jaworski-duck-sekaly,parfitt3,bronckers-etal}. In the model, \oca s represent single nucleated entities incorporated in a multinucleated active osteoclast, and \ocp s turn into \oca s upon \rankl-mediated activation of their $\rank$ receptor \cite{roodman,martin}. Transforming growth factor $\beta$ is known to be a general inhibitor of osteoclast differentiation and activation \cite{roodman}. For simplicity, here we only assume that \oca\ apoptosis is enhanced by the presence of \tgfb. Osteoclast maturation in the model can be summed up schematically as:
\begin{align}\label{oc}
    \ocp \stackrel{\rankl+}{\longrightarrow} \oca \stackrel{\tgfb+}{\longrightarrow} \emptyset.
\end{align}
We translate this sequence of events into the following balance equation for \oca s:
\begin{align}\label{oca}
    \frac{\p}{\p t} n_\oca = \mathcal{D}_\ocp({\footnotesize \rankl})\, n_\ocp - \mathcal{A}_\oca({\footnotesize \tgfb})\, n_\oca - \b\nabla\!\cdot\! \b J_\oca,
\end{align}
where $\mathcal{D}_\ocp$ is the \rankl-dependent differentiation rate of \ocp s and $\mathcal{A}_\oca$ the \tgfb-dependent apoptosis rate of \oca s. As in Ref.~\cite{pivonka-etal1}, the up-regulation and down-reglation of cellular responses by a ligand are assumed proportional to the fraction of occupied receptors. Mass action kinetics of the binding reactions shows that this is equivalent to modulating the cellular responses by certain ``activator'' and ``repressor'' functions of the ligand concentration (see Refs.~\cite{lemaire-etal,pivonka-etal1,lauffenburger-linderman} for more details). With the dimensionless activator and repressor functions
\begin{align}\label{piact-pirep}
    \piact(\xi) = \frac{\xi}{1+\xi}, \quad \pirep(\xi) = 1-\piact(\xi) = \frac{1}{1+\xi},
\end{align}
the functional forms of $\mathcal{D}_\ocp$ and $\mathcal{A}_\oca$ can thus be written as
\begin{align}
    &\mathcal{D}_\ocp({\footnotesize\rankl}) = D_\ocp\piact\Bigg(\frac{n_\rankl}{k^\rankl_\ocp}\Bigg),\label{D_OCp}
    \\&\mathcal{A}_\oca({\footnotesize\tgfb}) = A_\oca \piact\Bigg(\frac{n_\tgfb}{k^\tgfb_\oca}\Bigg),\label{A_OCa}
\end{align}
where $k^\rankl_\ocp$ and $k^\tgfb_\oca$ are dissociation binding constants, and $D_\ocp$ and $A_\oca$ are the maximal possible rates taken by $\mathcal{D}_\ocp$ and $\mathcal{A}_\oca$.

In the confined space of a cortical \bmu, cell diffusion is limited and we assume that directed motility dominates the movement of osteoclasts. The flux of \oca s can thus be written $\b J_\oca = n_\oca \b v_\oca$, where $\b v_\oca$ is the velocity of \oca\ cells with respect to the (fixed) bone matrix. The actual velocity of an active osteoclast is a combination of the dissolution process of the bone matrix, and of chemotactic and/or mechanotactic signals \cite{vaananen-etal,roodman,burger-etal,ishii-etal1,ishii-etal2}. Precisely how this sensing by osteoclasts of their mechanochemical micro-environment occurs is still uncertain and not an issue for the purposes of this paper. For this reason, in our model, the rate of movement of \oca s is simply taken to be constant, matching the average velocity $\b u$ of the \bmu's progression through bone:
\begin{align}
    \b J_\oca = n_\oca \b u. \label{oca-flux}
\end{align}
Note that typical cortical \bmu\ velocities range from 20 to 40~\um/\da\  \cite{parfitt3,martin-burr-sharkey,bone-mechanics-handbook}.

\subsubsection*{Osteoblasts}
Following the ODE model of bone remodelling proposed by Pivonka \etal\ \cite{pivonka-etal1}, three stages of osteoblast maturation are considered. ``Uncommitted progenitor osteoblasts'' (\obu s) denote a pool of mesenchymal stem cells assumed to be provided around the tip of the capillary \cite{jaworski-hooper,parfitt3,martin-burr-sharkey}. These cells are capable of committing to the osteoblastic lineage, becoming ``pre-osteoblasts'' (\obp s). This commitment is up-regulated by \tgfb\ \cite{harada-rodan,iqbal-sun-zaidi,tang-etal}. Pre-osteoblasts further mature into ``active osteoblasts'' (\oba s), found in large numbers (1000--2000 cells) at the back of cortical \bmu s (see Figure~\ref{fig:schematic-bmu}), actively laying down osteoid to refill the cavity opened by the osteoclasts \cite{parfitt3}. Based on Pivonka \etal~\cite{pivonka-etal1}, osteoblast activation is assumed to be down-regulated by \tgfb. The fate of active osteoblasts is either to be buried in osteoid and become osteocytes (approximately 95\% of all bone cells are osteocytes), to undergo apoptosis, or to become so-called bone-lining cells covering the surface of the new Haversian canal \cite{martin-burr-sharkey}. Elimination of \oba s from the active pool is assumed here to include all three possibilities. Osteoblast development in the model can thus be depicted as the sequence
\begin{align}\label{ob}
    \obu \stackrel{\tgfb+}{\longrightarrow} \obp \stackrel{\tgfb-}{\longrightarrow} \oba \longrightarrow \ldots,
\end{align}
leading to the following balance equations for \obp s and \oba s:
\begin{align}
    &\frac{\p}{\p t} n_\obp = \mathcal{D}_\obu({\footnotesize \tgfb})\, n_\obu - \mathcal{D}_\obp({\footnotesize \tgfb})\, n_\obp - \b\nabla\!\cdot\! \b J_\obp,\label{obp}
    \\&\frac{\p}{\p t} n_\oba = \mathcal{D}_\obp({\footnotesize \tgfb})\, n_\obp - A_\oba\, n_\oba - \b\nabla\!\cdot\! \b J_\oba,\label{oba}
\end{align}
where $\mathcal{D}_\obu, \mathcal{D}_\obp$ and $A_\oca$ are the \obu\ differentiation rate, the \obp\ differentation rate and the \oba\ elimination rate, respectively. Similarly to Eqs.~\eqref{D_OCp}--\eqref{A_OCa}, we set
\begin{align}
    &\mathcal{D}_\obu({\footnotesize \tgfb}) = D_\obu \piact\Bigg(\frac{n_\tgfb}{k^\tgfb_\obu}\Bigg), \label{D_OBu}
    \\&\mathcal{D}_\obp({\footnotesize \tgfb}) = D_\obp \pirep\Bigg(\frac{n_\tgfb}{k^\tgfb_\obp}\Bigg), \label{D_OBp}
\end{align}
with $k^\tgfb_\obu$, $k^\tgfb_\obp$ denoting dissociation binding constants and $D_\obu$, $D_\obp$ corresponding to the maximal possible rates taken by $\mathcal{D}_\obu$ and $\mathcal{D}_\obp$.

Active osteoblasts lay down osteoid in cortical \bmu s mainly radially, from the circumference of the cavity towards the center \cite{jaworski-hooper,parfitt3,martin-burr-sharkey}. As this process occurs on a time scale much larger than that of resorption, \oba s remain essentially stationary with respect to bone along the \bmu\ axis. Furthermore, it is observed that active osteoblasts, unlike osteoclasts, are not dynamically replenished once they have started producing osteoid \cite{parfitt3}. This suggests that the pre-osteoblasts they derive from are not moving longitudinally either (at least, not to a significant extent), and so we set $\b v_\obp = \b v_\oba =\b 0$, leading to
\begin{align} \label{ob-fluxes}
    \b J_\oba = \b J_\obp = \b 0.
\end{align}
As will be seen in Section \ref{sec:model-structure}, this hypothesis is crucial to explain the spatial segregation of active osteoblasts, pre-osteoblasts and uncommitted progenitors.

\subsubsection*{Regulatory factors and binding reactions}
System-level coupling between the osteoclasts and osteoblasts occurs because the two direct regulatory factors in our model (\tgfb\ and \rankl; see Eqs.~\eqref{oc} and \eqref{ob}) are themselves driven by the cellular actions, both directly and indirectly via other interfering molecules.

\Tgfb\ is stored in high concentration in the bone matrix and released into the \bmu\ environment in active form by the resorbing osteoclasts \cite{roodman,iqbal-sun-zaidi,tang-etal}. Assuming that \tgfb\ degrades at a constant rate $D_\tgfb$, we have:
\begin{align}\label{tgfb}
    \frac{\p}{\p t}n_\tgfb = n^\text{bone}_\tgfb \kres n_\oca - D_\tgfb n_\tgfb - \b\nabla\!\cdot\!\b J_\tgfb,
\end{align}
where $\kres$ is the bone volume resorbed per unit time by a single osteoclast and $n^\text{bone}_\tgfb$ is the concentration of \tgfb\ present in the bone matrix. Since \tgfb\ is released in the environment in soluble form, its transport properties encoded in $\b J_\tgfb$ are assumed to be independent of the cells. It is expected that high levels of \tgfb\ are found up until the reversal zone where it promotes commitment and differentiation of mesenchymal cells to the osteoblastic lineage. For simplicity, we assume that \tgfb\ has negligible diffusion, \ie, $\b J_\tgfb \approx \b 0$. Nevertheless, the presence of \tgfb\ in the reversal zone can be accounted for by assuming a weak degradation rate $D_\tgfb$ (in a sense clarified below). Further comments on the effects of \tgfb\ diffusion towards the back of the \bmu\ are made in Section~\ref{sec:model-structure}.

The local availability of \rankl, which is critical for the differentiation of \ocp s into \oca s, arises from the combination of several effects. \Rankl\ is a protein bound to the membrane of cells of the osteoblastic lineage. Its interaction with the \rank\ receptor found on \ocp\ is regulated by the presence of the soluble decoy receptor \opg, which is also expressed by osteoblastic cells. Furthermore, the relative expression of \rankl\ \vs\ \opg\ by osteoblasts is regulated by systemic \pth\ concentrations. All these molecules and their competitive interactions are considered explicitly in our model. Here we only describe their main features, and refer the reader to Ref.~\cite{pivonka-etal1} for further details. We will assume that \rankl\ is only expressed by \obp s and that \opg\ is only expressed by \oba s (corresponding to ``Model Structure 2'' of Ref.~\cite{pivonka-etal1}). This choice of ligand expression is in agreement with experimental findings \cite{gori-etal,thomas-etal} and the conclusions drawn in Ref.~\cite{pivonka-etal1}. However, to reexamine this assumption in a spatio-temporal framework, we will study its influence in Section~\ref{sec:model-structure}. While the flux of soluble \opg\ is assumed independent of the cells (similarly to \tgfb), transport of membrane-bound \rankl\ is tied to the cells expressing it: $\b J_\rankl = n_\rankl \b v_\ob$. However, osteoblasts are assumed to have negligible motility ($\b v_\ob\approx \b 0$), and so $\b J_\rankl \approx \b 0$.

A considerable simplification of the mass action kinetic equations considered for the competitive bindings between \rank, \rankl\ and \opg\ was obtained in Ref.~\cite{pivonka-etal1} due to the separation of time scales between the fast reaction rates of ligands binding to their receptors on cells, and comparatively slow cell responses. We examine here the consequence of this separation of time scales in the presence of transport terms in Eq.~\eqref{mass-balance}. Let $r_L$ be the slowest reaction rate (\eg, in $\da^{-1}$) to be found in the source/sink terms in $\sigma_L$ for the ligand $L$. Dividing Eq.~\eqref{mass-balance} by $r_L$, one has
\begin{align}\label{pre-fast-binding}
    r_L^{-1} \frac{\p}{\p t} n_L = r_L^{-1} \sigma_L - r_L^{-1}\b\nabla\!\cdot\!\b J_L
\end{align}
If reaction binding dominates transport, then \mbox{$|r_L^{-1}\b\nabla\!\cdot\!\b J_L|\ll 1$} and $r_L^{-1}\sigma_L=\Order(1)$. Thus, changes in the local concentration of the free ligand occur on the short timescale $r_L^{-1}$ and only quasi-steady states need to be considered for the cellular dynamics, leading to
\begin{align}\label{fast-binding}
    \sigma_L \approx 0\qquad\forall \b r, t.
\end{align}
This simplification is exactly of the same form as in the temporal model \cite[Eqs.~(16)--(20)]{pivonka-etal1}. We assume here that it holds for \rankl, \opg\ and for \pth. As in Ref.~\cite{pivonka-etal1}, Eq.~\eqref{fast-binding} is thus used to express the concentrations of these regulatory factors in terms of the remaining unknowns of the system. This has been presented in detail in Ref.~\cite{pivonka-etal1}, so we only briefly mention the results here. The \pth\ endogeneous production rate  $P_\pth(\b r,t)$ and degradation rate $D_\pth$ are assumed to be given and not further regulated. Thus, Eq.~\eqref{fast-binding} with $\sigma_\pth(\b r,t) = P_\pth-D_\pth n_\pth$ determines the \pth\ concentration as
\begin{align}
n_\pth=P_\pth/D_\pth \label{pth}
\end{align}
(see Eq. (25) of \cite{pivonka-etal1}). Production and elimination rates of \rankl\ and \opg\ in Ref.~\cite{pivonka-etal1} have a more complicated form owing to their regulation by \pth, the interdependence between \rank, \rankl\ and \opg, and an assumed saturation of the endogeneous production responses. With similar notations as in Ref.~\cite[Eqs.~(30)--(36)]{pivonka-etal1}, the concentrations of \opg\ and \rankl\ can be rewritten with the help of the functions~\eqref{piact-pirep} as:
\begin{align}
    &n_\opg = \opg_\text{max} \piact\Bigg( \frac{\beta_{1,\opg} n_\obp+\beta_{2,\opg} n_\oba}{\opg_\text{max} D_\opg} \pi^\pth_{\text{rep},\ob}\Bigg),
    \\&n_\rankl=\frac{\beta_\rankl}{D_\rankl} \pirep\left(k_{A1,\rankl} n_\opg+k_{A2,\rankl} n_\rank\right)\notag \\&\times\piact\Bigg(\frac{D_\rankl}{\beta_\rankl}(R^\rankl_1 n_\obp\!+\!R^\rankl_2 n_\oba) \pi^\pth_{\text{act},\ob}\Bigg),\label{rankl}
\end{align}
In Ref.~\cite{pivonka-etal1} and in the present model, the same constant number of \rank\ receptors $N^\rank_\ocp$ is assumed to be expressed on each \ocp\ cell. However, while the density of \ocp s was constant in Ref.~\cite{pivonka-etal1}, it is space and time dependent here. The constant \rank\ concentration occurring in the Eq.~(36) of Ref.~\cite{pivonka-etal1} has thus to be replaced in Eq.~\eqref{rankl} above by the local, time-dependent concentration
\begin{align}
    n_\rank = N^\rank_\ocp n_\ocp. \label{rank}
\end{align}
Unlike Ref.~\cite{pivonka-etal1}, we do not assume that Eq.~\eqref{fast-binding} holds for \tgfb, and keep its differential description given by Eq.~\eqref{tgfb}. Indeed, the production rate of \tgfb\ occurs on a cellular time scale and its degradation rate is assumed to match the slower characteristic times of the cellular dynamics.

We finally note that Eqs.~\eqref{pre-fast-binding}--\eqref{fast-binding} also apply to the balance of bound receptor--ligand complexes. Their fast binding properties allow us to express via \eqref{fast-binding} the receptor occupancy per cell in terms of the concentration of free ligand as has been used in Eqs.~\eqref{D_OCp}, \eqref{A_OCa}, \eqref{D_OBu}, \eqref{D_OBp} with the functions \eqref{piact-pirep}.

\subsubsection*{External conditions}
Because all cells eventually differentiate further or undergo apoptosis, a continual supply of precursor cells is needed to reach nonzero cell populations over a period of time exceeding a couple of days. In cortical remodelling, this supply is local: it reaches the reversal zone of the \bmu\ through an internal capillary that grows at the same rate as the \bmu\ progresses (see Figure \ref{fig:schematic-bmu}) \cite{parfitt3}. We assume here that the replenishment of \obu\ and \ocp\ cells occurs around the tip of the capillary in an unbounded and non-rate-limiting way. Under that assumption, the inhomogeneous densities $n_\ocp$ and $n_\obu$ instantaneously reach a stationary distribution peaked around the tip of the growing capillary \cite{parfitt3}. These densities become given external functions in Eqs.~\eqref{oca} and \eqref{obp}, of the form
\begin{align}
    n_\ocp(\b r, t) = \ocp(\b r\!-\!\b u t),\quad n_\obu(\b r, t) = \obu(\b r \!-\!\b u t).
\end{align}
We assume $\ocp(\b r)$ and $\obu(\b r)$ to be Gaussian distributions centered around the capillary tip (see Figure~\ref{fig:bmu-evolution}).

Finally, while \pth\ has been included into the model following Ref.~\cite{pivonka-etal1}, its spatial implications in the \bmu\ will not be investigated for the purpose of the present study, and we assume that the concentration of \pth\ is constant and homogeneously distributed along the \bmu.

Solving the system of PDEs \eqref{oca}, \eqref{obp}, \eqref{oba} and \eqref{tgfb} requires appropriate initial and boundary conditions. In the following, these equations are solved in one spatial dimension with boundary conditions specified at the very front of the \bmu\ and at its back.

\section{Density profiles in the \bmu}\label{sec:density-profiles}
As spatial profiles in a \bmu\ are predominantly structured along the longitudinal $x$-axis (see Figure~\ref{fig:schematic-bmu}), we restrict the mathematical model to this single spatial dimension, neglecting variations in transverse cross-sections: $n_A(\b r,t) \approx n_A(x,t)$. As explained in Section~\ref{sec:model}, the fast binding approximation \eqref{fast-binding} allows to substitute $n_\pth$, $n_\rank$, $n_\rankl$, and $n_\opg$ with their expression~\eqref{pth}--\eqref{rankl} in the PDEs \eqref{oca}, \eqref{obp}, \eqref{oba} and \eqref{tgfb}, which are then solved numerically (using \texttt{Mathematica} \cite{mathematica}) for the remaining unknown concentration profiles $n_\oca$, $n_\obp$, $n_\oba$ and $n_\tgfb$. These PDEs are of the reaction-advection type and a boundary condition needs to be specified at a single point of the $x$ axis in each equation. Based on bone physiology, we prescribe both $n_\obp$ and $n_\oba$ to be zero at the tip of the \bmu\ cavity, and both $n_\oca$\ and $n_\tgfb$ to be zero at the back of the \bmu, where the new osteon cavity is refilled with osteoid up to the diameter of the Haversian canal. These requirements in turn specify a \bmu\ spatial domain over which the PDEs are solved. This domain is set on either side of the capillary tip (which moves along $x$ at rate $u$) as follows. The tip of the \bmu\ cavity is defined to be 350~\um\ ahead of the capillary tip while the back of the \bmu\ is defined to be 4800~\um\ behind the capillary tip \cite{martin-burr-sharkey}, thus allowing the \bmu\ to spread over about 5~\mm.\footnote{Note, however, that cell densities can be concentrated on a more restricted portion of this domain.}$^,$\footnote{The origin of the (static) $x$-axis is also chosen such that it coincides with the tip of the moving cavity at time $t=100$~\da s (see Figure \ref{fig:bmu-evolution}).} To transform the moving-boundary conditions into time-independent conditions, the problem is solved in a reference frame co-moving with the \bmu\ at rate $u$ along $x$ (see Ref.~\cite{buenzli-pivonka-etal} for more details).

\subsection{Numerical results and discussion}\label{sec:results-discussion}

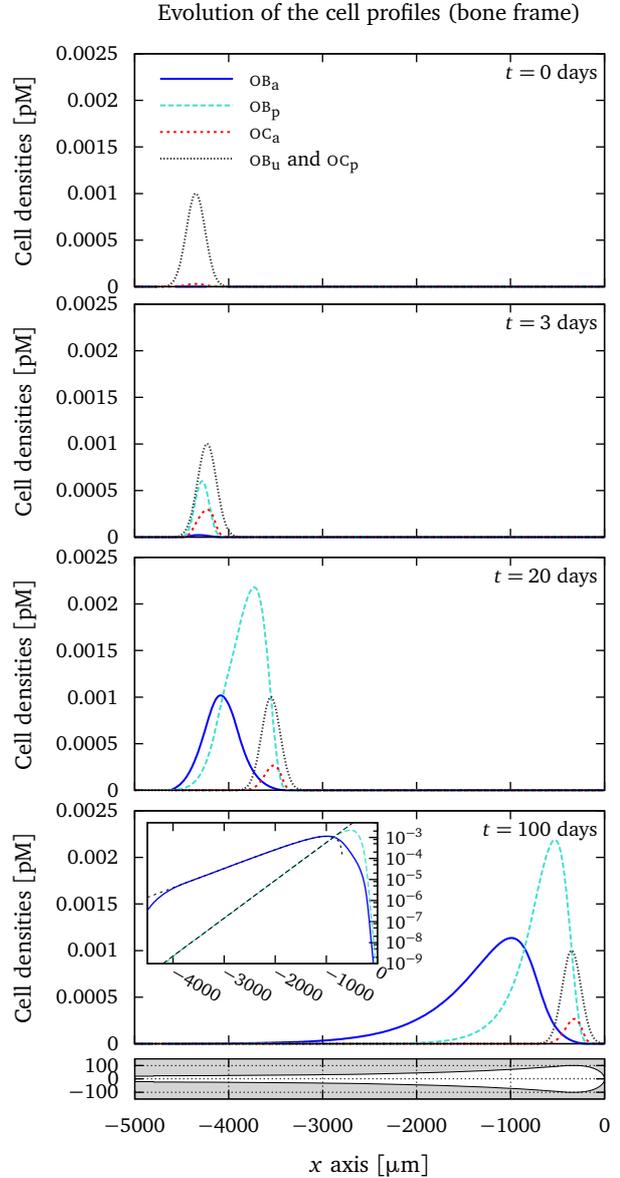
\begin{figure}
        \centering{%
        \makebox[3.083in]{\input{profiles-modified-root-model-bone-frame-at-0d}}
        \vskip1mm
        \makebox[3.083in]{\input{profiles-modified-root-model-bone-frame-at-3d}}
        \vskip1mm
        \makebox[3.083in]{\input{profiles-modified-root-model-bone-frame-at-20d}}
        \vskip1mm%
        \makebox[3.083in]{\input{profiles-modified-root-model-bone-frame-at-100d-with-inset}}
        \vskip0.4mm%
        \makebox[3.083in]{\input{bmu-cavity-sketch-bone-frame-at-100d}}%
        }%
    \caption{Evolution of the cell profiles computed from the mathematical model. At $t=0$ days, the initial conditions are shown: only precursor cells and a tiny population of \oca s are present. At $t=3$ days, these initial \oca s have released enough \tgfb\ in the environment to trigger differentiation of \obu s into \obp s, which in turn have increased the \oca\ population by \rankl-binding. At $t=20$ days, the profiles at the front of the \bmu\ have already taken a constant shape. Active osteoblasts have started to emerge behind \obp s. At $t=100$ days, all profiles have essentially reached a quasi-steady-state: they are moving forward into the bone matrix without changing shape. A sketch of a typical \bmu\ cavity is aligned with these steady-state profiles for comparison. \textit{Inset:} \obp\ and \oba\ profiles represented in logarithmic scale together with the asymptotic expressions given in Eqs.~\eqref{obp-oba-asymptotics} ($a=0.036$, $b=-0.0545$).}
    \label{fig:bmu-evolution}
\end{figure}

The evolution of the computed cell profiles is shown from the (static) bone frame in a series of temporal snapshots in Figure~\ref{fig:bmu-evolution}. These profiles define the shape of a multi-cellular wave front emerging and propagating into the bone at constant velocity $u=40~\um/\da$, corresponding to a remodelling \bmu. Starting at $t=0$ days from a tiny population of active osteoclasts present around the precursor cells (assumed to be recruited there during an ``activation'' phase of the \bmu), the densities of \obp\ and \oca\ cells quickly increase to reach quasi-steady profiles at the front of the \bmu\ over the next $20$ days, progressing forward without changing shape. The tails of the \obp\ and (particularly) \oba\ profiles further at the back, however, take longer to develop. As a result of the differentiation sequences \eqref{oc} and \eqref{ob} the heights of the profiles automatically reach bounded values after an establishment phase and they are confined over a spatial range corresponding to the known dimensions of a \bmu\ (of the order of a few millimetres). The development of a clearly structured travelling wave profile is predicted by the model, as is observed histologically \cite{parfitt3,martin-burr-sharkey,bone-mechanics-handbook}. Pre-osteoblasts and active osteoblasts are shifted towards the back of the \bmu. The inversion of the relative number of \obp s \textit{vs} \oba s at around $-850~\um$ (at $t=100$~days)  in Figure~\ref{fig:bmu-evolution} shows that the model captures the transition between the reversal zone and the formation zone along the longitudinal axis of the \bmu\ (compare also with Figure~\ref{fig:schematic-bmu}).

On the other hand, it is apparent that the model does not capture the transition between the resorption zone and the reversal zone: the populations of \oca s and \ocp s overlap everywhere at the front of the \bmu. The bone remodelling biochemistry implicated in the model thus far, whilst reproducing bone-formation features of the \bmu\ very well, is not satisfactory in explaining the spatial cell structure in the resorption zone, which suggests there are missing biochemical components not taken into account in this first model. In Section~\ref{sec:model-structure}, we therefore supplement our model with a further cellular component to resolve this behaviour.

\subsection{Theoretical analysis of the cell profiles}
The simple wave-form Ansatz $n_A(x,t) \equiv A(x-u t)$ for the density profiles reduces the system of PDEs to a system of ordinary differential-algebraic equations (DAEs) to solve for the steady-state profiles $A(x)$, $A=\obp,\oba,\oca,\tgfb$ (see also Ref.~\cite{buenzli-pivonka-etal}).\footnote{The equation for $\oca(x)$ becomes algebraical while the other equations keep a differential nature.} For \obp\ and \oba, these equations are
\begin{align}
    &-u\, \frac{\p}{\p x}\obp = \mathcal{D}_\obu({\footnotesize \tgfb}) \obu - \mathcal{D}_\obp({\footnotesize \tgfb}) \obp, \notag
    \\&-u\, \frac{\p}{\p x}\oba = \mathcal{D}_\obp({\footnotesize \tgfb}) \obp - A_\oba \oba.
    \label{obp-oba-eq-steady}
\end{align}
We can use Eqs.~\eqref{obp-oba-eq-steady} to quantify the shifts of the osteoblastic profiles towards the back of the \bmu\ as well as their relative height in terms of model parameters linked to biological characteristics of the system. Let $x_\obp^\ast$ and $x_\oba^\ast$ be the positions of the maximum in the \obp\ and \oba\ steady-state profiles (in Figure \ref{fig:bmu-evolution}, these are located at $x_\obp^\ast\approx -550~\um$ and $x_\oba^\ast\approx -1000~\um$). Since the spatial derivative of $\obp(x)$ vanishes at $x_\obp^\ast$, and that of $\oba(x)$ at $x_\oba^\ast$, one obtains from \eqref{obp-oba-eq-steady} the following ratios of the densities of $\obp$ \vs\ $\obu$ and the densities of $\oba$ \vs\ $\obp$ at these points:
\begin{align}
    &\frac{\obp}{\obu}(x_\obp^\ast) = \frac{D_\obu}{D_\obp} f\big({\footnotesize \tgfb(x_\obp^\ast)}\big), \notag
    \\&\frac{\oba}{\obp}(x_\oba^\ast) = \frac{D_\obp}{A_\oba} g\big({\footnotesize \tgfb(x_\oba^\ast)}\big), \label{ob-ratios}
\end{align}
where
\begin{align}
&f\big({\footnotesize \tgfb}\big)=\piact\Big(\tfrac{\tgfb}{k^\tgfb_\obu}\Big)/\pirep\Big(\tfrac{\tgfb}{k^\tgfb_\obp}\Big), \notag
\\&g\big({\footnotesize \tgfb}\big) = \pirep\Big(\tfrac{\tgfb}{k^\tgfb_\obp}\Big)
\end{align}
are monotonously increasing and decreasing functions of \tgfb, respectively. Due to the couplings existing between the various regulatory factors and the cells in the model, the \tgfb\ concentration occurring in the right hand side of Eqs.~\eqref{ob-ratios} depends implicitly on all cell densities (and in particular on their ratios \obp/\obu, \oba/\obp), so Eqs.~\eqref{ob-ratios} do not express the cell density ratios in the left hand sides in closed form, and other dependences upon the fractions $D_\obu/D_\obp$ and $D_\obp/A_\oba$ exist in $f$ and $g$, respectively. Nevertheless, it can be checked numerically that the parameter fractions $D_\obu/D_\obp$ and $D_\obp/A_\oba$ are main regulators of the cell density ratios in the left hand side of Eqs.~\eqref{ob-ratios}. In fact, it can be argued that the implicit dependence of $f$ and $g$ on these parameter fractions enhances the explicit linear dependences seen in Eqs.~\eqref{ob-ratios}. Indeed, upon increasing $D_\obu/D_\obp$, \rankl\ is increased over \opg, leading to an increase of \oca, thus of \tgfb\ and of $f$. On the other hand, upon increasing $D_\obp/A_\oba$, \opg\ is increased over \rankl, leading to a decrease of \oca\ and of \tgfb, thus to an increase of~$g$.

Multiplying the second equation in \eqref{obp-oba-eq-steady} by $1$ or by $x$ and integrating it over the length of the steady-state \bmu\ (from $-\infty$ to $0$), one can use integration by parts, the boundary condition $\oba(0)=0$ and the fact that $\oba(x)$ decays exponentially as $x\to-\infty$ (see Eq. \eqref{obp-oba-asymptotics}) to derive the following relations:
\begin{align}
    &0=\int_{-\infty}^0 \!\!\!\!\!\!\d x\ \mathcal{D}_\obp \obp - A_\oba \!\!\int_{-\infty}^0 \!\!\!\!\!\!\d x\ \oba, \label{ob-integrals}
    \\&u \!\int_{-\infty}^0 \!\!\!\!\!\!\d x\ \oba = \int_{-\infty}^0 \!\!\!\!\!\!\d x\ x\, \mathcal{D}_\obp \obp - A_\oba \!\!\int_{-\infty}^0 \!\!\!\!\!\!\d x\ x\, \oba. \notag
\end{align}
Under the assumption that the variation of $\mathcal{D}_\obp$ along $x$ can be neglected for the values of the integrals, one can factor $\mathcal{D}_\obp$ out of the integrals.\footnote{In fact, $\mathcal{D}_\obp$ and $\mathcal{D}_\obu$ vary significantly over the domain of integration. However, it is possible to use a generalised integral mean-value theorem (see \cite[\S 27]{halmos}) to factor these functions out of the integrals. This leads to exact relations between $\avg{x_\oba}$ and $\avg{x_\obp}$ and between $\avg{x_\obp}$ and $\avg{x_\obu}$ similar to Eqs.~\eqref{oba-shift}--\eqref{obp-shift}.} Given that the average position of the profile $A(x)$ (or ``center of mass'' along $x$) is $\avg{x_A} \equiv \int \!\d x\, x A(x)/\int \!\d x\, A(x)$, one thereby obtains from Eqs.~\eqref{ob-integrals}:
\begin{align}
    \avg{x_\oba} \approx \avg{x_\obp} - \frac{u}{A_\oba}.\label{oba-shift}
\end{align}
Eq.~\eqref{oba-shift} now shows that the ratio $u/A_\oba$ is essentially determining the length of the shift of the \oba\ profile towards the back of the \bmu\ compared to \obp s. Performing a similar analysis with the first equation in \eqref{obp-oba-eq-steady}, one has
\begin{align}
    \avg{x_\obp} \approx \avg{x_\obu} - \frac{u}{\mathcal{D}_\obp},\label{obp-shift}
\end{align}
meaning that \obp s are shifted to the back of \obu s by a length proportional to $u/D_\obp$.

Finally, it is possible to give analytical expressions for the decays of the \obp\ and \oba\ profiles at the far back of the \bmu. Indeed, in this region, \tgfb\ is essentially absent, so that $\mathcal{D}_\obu \approx 0$, and $\mathcal{D}_\obp \approx D_\obp$ (see Eqs.~\eqref{D_OBu}, \eqref{D_OBp}). Eqs. \eqref{obp-oba-eq-steady} thus become linear and their solution can be calculated explicitly, leading to the asymptotic behaviours
\begin{align}
    &\obp(x) \sim a\ \e^{-\frac{D_\obp}{u} |x|}, \label{obp-oba-asymptotics}
    \\&\oba(x) \sim b\ \e^{-\frac{A_\oba}{u} |x|} + \frac{a}{1-\frac{A_\oba}{D_\obp}} \Big[ \e^{-\frac{A_\oba}{u}|x|} - \e^{-\frac{D_\obp}{u}|x|} \Big] \notag
\end{align}
as $x\to-\infty$, where $a$ and $b$ are two integration constants. These asymptotic behaviours of the steady-state profiles are compared with the numerical profiles obtained by the temporal evolution in Figure \ref{fig:bmu-evolution} in logarithmic scale. While the slope of these curves is essentially determined by the ratios $D_\obp/u$ and $A_\oba/u$, the constants $a$ and $b$ shift the height of the curves in the logarithmic plot, and they have been fitted. The small discrepancy visible at the very back of the \bmu\ is due to the fact that the numerical profiles have not yet reached a quasi-steady state there.

Equations \eqref{ob-ratios}, \eqref{oba-shift}, \eqref{obp-shift} and \eqref{obp-oba-asymptotics} all relate kinetic properties of the cells (velocity, differentiation and apoptosis rates) to intrinsic features of spatial profiles in a cortical \bmu\ at a fixed snapshot in time. This entanglement of time and space reflects the wave-like character of the \bmu's progression. Importantly, it is noted that the experimental observation of such profiles from histological analyses could allow a direct estimation of the cellular kinetic properties in this model.

\section{Roles of model assumptions and extension of the model}\label{sec:model-structure}
In this section, we use our mathematical model to examine the influence and assess the validity of several assumptions made in Section~\ref{sec:model}. Based on the discussion in Section~\ref{sec:results-discussion} and the identified model shortcoming, we then extend the model by adding a new stage of osteoclast differentiation. This extension resolves the inability of the previous model to capture the transition between the resorption zone and the reversal zone.

\subsubsection*{Influence of cell motility}
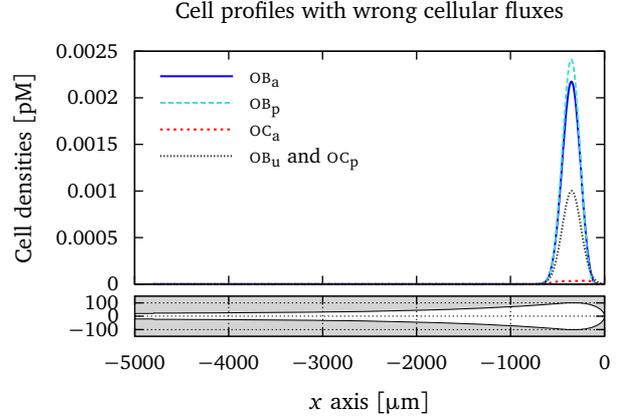
\begin{figure}
        \centering{%
        \makebox[3.083in]{\input{profiles-modified-root-model-allFluxesnv-at-100d}}
        \vskip0.4mm%
        \makebox[3.083in]{\input{bmu-cavity-sketch-bone-frame-at-100d}}%
        }%
    \caption{Steady-state density profiles obtained when all cell velocities are taken equal, \ie, $v_\obp=v_\oba = v_\oca = u$. All parameters are otherwise taken as in Figure \ref{fig:bmu-evolution}. The typical spatial organisation of the bone cells in a \bmu\ is not reproduced in this case. These profiles are in clear mismatch with the typical shape of the \bmu\ cavity sketched below.}
    \label{fig:all-fluxes-nv}
\end{figure}

While the effects of many regulatory factors on cell commitment, proliferation, differentiation and apoptosis are well-established in the context of bone remodelling, less is known on the motile properties of the bone cells and the regulation thereof, although recent progress has been made in this direction \cite{tang-etal,ishii-etal1,ishii-etal2}.

Here we show that these motile properties can influence dramatically the spatio-temporal coordination of the bone cells, and thus the functioning of bone remodelling. In Section \ref{sec:model}, \obp s and \oba s were assumed to stay stationary with respect to bone and we set their velocities to zero (see Eqs.~\eqref{ob-fluxes}). The wave-like propagation of the density of \obp\ and \oba\ cells at rate $u$ observed in Figure \ref{fig:bmu-evolution} is entirely due to their creation upstream and elimination downstream. On the other hand, choosing \obp\ and \oba\ cell velocity to correspond to $u$, so that $v_\obp = v_\oba = v_\oca = u$, still leads to a wave-like propagation of the cell densities at rate $u$. However, as can be seen from Figure~\ref{fig:all-fluxes-nv}, in this situation all the cell density profiles fall within the same region around the precursor cell source. Such cellular profiles are in clear mismatch with the experimentally-observed propagation of a structured cortical \bmu\ with a separation in time and in space of the resorption and formation processes. Since all cells and regulatory factors overlap, their net interaction is modified and the cell density profiles also reach different heights, offsetting the bone balance. This result corroborates the experimental observation that osteoblasts do not move significantly after their commitment to the osteoblastic lineage \cite{parfitt3} and emphasises the critical role that cell motility plays in \bmu-based remodelling.

\subsubsection*{Role of osteoblastic maturation stage for expression of \rankl\ and \opg}
\begin{figure}
        \centering{%
        \makebox[3.083in]{\input{profiles-modified-root-model-OpgRanklSwap-at-100d}}
        \vskip0.4mm%
        \makebox[3.083in]{\input{bmu-cavity-sketch-bone-frame-at-100d}}%
        }%
    \caption{Steady-state density profiles obtained when the expression of \rankl\ and \opg\ on precursor and mature osteoblasts is swapped, \ie, \rankl\ is expressed on \oba s and \opg\ is expressed on \obp. All parameters are otherwise taken as in Figure \ref{fig:bmu-evolution}. The density of active osteoclasts has not grown past its small initial condition. Such a small population of \oca s would not be able to create a \bmu\ cavity as before.}
    \label{fig:opg-rankl-swap}
\end{figure}
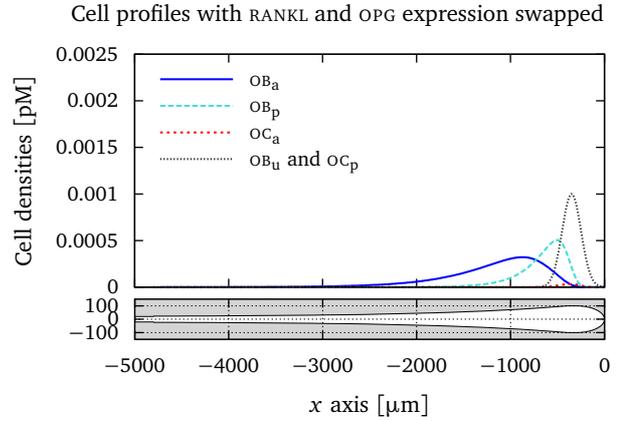
Several experiments have characterised \rankl\ and \opg\ expression levels on maturing osteoblasts, concluding that \rankl\ expression dominates in immature osteoblasts while \opg\ expression dominates in more mature osteoblasts \cite{gori-etal,thomas-etal}. In the purely temporal model of Pivonka \etal\ \cite{pivonka-etal1}, various ``model structures'' of expression of \rankl\ and \opg\ by osteoblasts at different stages of maturation were tested, which supported these experimental findings. However, these model structures really need to be tested for their functional importance in the context of the spatio-temporal coordination of bone cells in a \bmu. The density profiles predicted by our model when \rankl\ is only expressed on \oba s and \opg\ is only expressed on \obp s (corresponding to ``Model Structure 1'' of Ref.~\cite{pivonka-etal1}) are shown in Figure \ref{fig:opg-rankl-swap}. Clearly, the lack of availability of \rankl\ in the reversal zone, due to its expression on \oba s at the back of the \bmu\ and its further screening by \opg\ expressed on \obp s, blunts activation of osteoclasts. Barely any \oca s are found in the steady-state. Such a tiny population of \oca s would not be able to create a \bmu\ cavity with a size comparable to experimentally-observed \bmu s. These results thus strongly support the model structure where \rankl\ is expressed on \obp s and \opg\ is expressed on \oba s (\ie, ``Model Structure 2'' of Ref.~\cite{pivonka-etal1}), which is used throughout the paper.

\subsubsection*{Role of {\tgfb}}
\Tgfb\ is a multi-facetted regulatory factor serving several purposes in bone remodelling \cite{iqbal-sun-zaidi,tang-etal}. In Ref.~\cite{pivonka-etal1} and in our model, \tgfb\ up-regulates osteoblast commitment but down-regulates osteoblast activation (see Eq.~\eqref{ob}). Furthermore, it up-regulates active osteoclast apoptosis. These several roles find sense in the structured cell profiles of a cortical \bmu. Physiologically, a negative feedback loop on osteoclasts is needed to keep resorption under control. Having \tgfb\ regulating this negative feedback is convenient, since it is then activated only once bone has started to be resorbed. On the other hand, the portion of bone just resorbed needs to be refilled with new bone. While \tgfb\ diffuses behind \oca s in the cortical \bmu, it reaches the reversal zone populated with \obu s. \Tgfb\ commands their commitment to the refilling task by up-regulating their development into \obp s. Finally, the down-regulation of activation of \obp\ into \oba\ by \tgfb\ helps preventing osteoid to be laid down too early, \eg\ in the resorption zone. From our model, we observe that the presence of \tgfb\ behind \oca s acts to delay the onset of \oba s and so to increase the spatial segregation between \oca s and \oba s.

\subsubsection*{Model extension: including ``mature osteoclasts''}
Precursor osteoclasts are known to be circulating cells \cite{martin,roodman} and delivered to the reversal zone of cortical \bmu s through the capillary (see Figure~\ref{fig:schematic-bmu}). Throughout the \bmu's progression, the capillary tip is found at a distance of about $350~\um$ behind the resorbing front. This means osteoclasts need to travel this distance at a faster pace than the rate of progression $u$ of the \bmu\ to reach the front \cite{parfitt3}. On the other hand, activation of osteoclasts requires \rankl, which is expressed on the surface of pre-osteoblasts found around the capillary tip. In the model presented in Section~\ref{sec:model}, \oca s are assumed to resorb the bone matrix, so while they have been activated by \rankl, they cannot move faster than $u$. For this reason, \oca s are differentiating from \ocp s around the middle of the reversal zone and stay there as they have not been given regulatory mechanisms to distance themselves from their progenitors (such as chemotactic signals towards the bone surface \cite{ishii-etal2}). This results in overlapping \ocp\ and \oca\ density profiles in Figure~\ref{fig:bmu-evolution}.

\begin{figure}
        \centering{%
        \makebox[3.083in]{\input{profiles-model2-at-100d-cells}}
        \vskip0mm
        \makebox[3.083in]{\input{profiles-model2-at-100d-factors}}
        \vskip0.4mm%
        \makebox[3.083in]{\input{bmu-cavity-sketch-to-3000}}%
        }%
    \caption{Cell density and regulatory concentration profiles in a cortical \bmu\ as given by the extended model (compare with Figure \ref{fig:bmu-evolution}). Active osteoclasts are now clearly shifted towards the front of the \bmu\ compared to their progenitors. The overlap between \oba\ and \oca\ is practically inexistant. The osteoblastic profiles in the back are essentially unchanged except for their amplitude. The concentrations of \opg, \rankl\ and \tgfb\ are also shown. They have been scaled to correspond to \oba, \obp\ and \oca\ density levels, respectively, for comparison. Remarkably, \obp-bound free \rankl\ is not colinear with the presence of \obp: it is effectively shifted towards osteoclast precursor cells due to the presence of \opg. The concentration of \tgfb\ is also found behind the \oca\ peak.}
    \label{fig:bmu-model-2}
\end{figure}
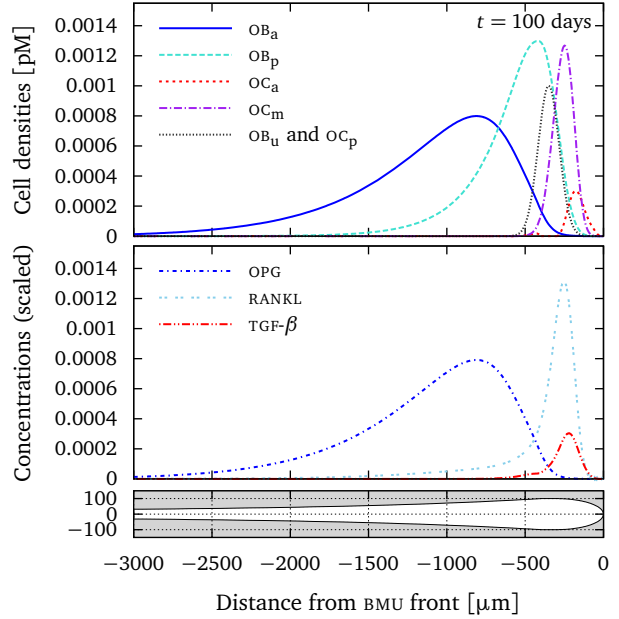
To resolve this problem, we are led to introduce a new stage of osteoclast development in the model, that we call ``mature osteoclasts'' (\ocm s). Mature osteoclasts denote osteoclasts that have been activated by \rankl\ and that migrate towards the front at a velocity $v_\ocm>u$ until they reach the bone surface. Once at the bone surface, they join an active resorbing multinucleated osteoclast, hence becoming \oca s progressing at rate $u$. The sequence of osteoclast maturation in Eq.~\eqref{oc} is thereby extended to
\begin{align}\label{oc-extended}
    \ocp \stackrel{\rankl+}{\longrightarrow} \ocm \stackrel{\text{bone surface}}{\longrightarrow} \oca \stackrel{\tgfb+}{\longrightarrow} \emptyset.
\end{align}

To model the transition from \ocm\ to \oca\ at the bone surface of the genuinely three-dimensional \bmu\ cavity in our one-dimensional setup, we introduce a ``probability of reaching bone'' $\Phi(x,t)$, defined as
\begin{align}
    \Phi(x,t) = 1-\frac{R^2(x,t)}{R_c^2},\label{Phi}
\end{align}
where $R(x,t)$ is the radius of the \bmu\ cavity depicted in Figures \ref{fig:bmu-evolution}--\ref{fig:bmu-model-2}, and $R_c=100~\um$ is the maximal cavity radius (cement line radius).\footnote{This probability corresponds to the bone volume fraction of a cross-sectional slice at $x$ of a rotationally-symmetric cortical \bmu\ of radius $R(x,t)$ compared to the cylinder of radius $R_c$.} For simplicity, this cavity is assumed here to be given and to progress forward at rate $u$ without changing shape (an assumption valid in the quasi-steady-state). According to Eq.~\eqref{Phi}, the function $\Phi(x,t)$ thus interpolates between one ahead of the (moving) cavity front (where the cavity radius $R(x,t)$ is formally zero), and zero in the reversal zone (where $R(x,t)$ reaches the cement line radius $R_c$). The function $\Phi(x,t)$ increases again to a positive value $<1$ for $x$ in the formation zone (where $R(x,t)$ decreases towards the Haversian canal radius). With this definition, the sequence of osteoclast development in Eq.~\eqref{oc-extended} can be expressed as:
\begin{align}
    &\frac{\p}{\p t} n_\ocm = \mathcal{D}_\ocp({\footnotesize \rankl})\, n_\ocp - \mathcal{D}_\ocm({\footnotesize \Phi})\, n_\ocm - \b\nabla\!\cdot\! \b J_\ocm,\notag
    \\&\frac{\p}{\p t} n_\oca = \mathcal{D}_\ocm({\footnotesize \Phi})\, n_\ocm - \mathcal{A}_\oca({\footnotesize \tgfb})\, n_\oca - \b\nabla\!\cdot\! \b J_\oca,\label{ocm-oca}
\end{align}
where $\mathcal{D}_\ocp$, $\mathcal{A}_\oca$ and $\mathcal{\b J_\oca}$ are given by Eqs.~\eqref{D_OCp}, \eqref{A_OCa} and~\eqref{oca-flux}. The transition from \ocm\ to \oca\ is assumed to take place at a rate $\mathcal{D}_\ocm$ proportional to the probability of reaching bone:
\begin{align}
    &\mathcal{D}_\ocm({\footnotesize \Phi}) = D_\ocm \Phi,\label{D_OCm}
\end{align}
where $D_\ocm$ is the rate of osteoclast activation once at the bone surface. Finally, the flux of \ocm s  is taken to be
\begin{align}
    &\b J_\ocm = n_\ocm \b v_\ocm. \label{J_OCm}
\end{align}
where the velocity is such that $v_\ocm>u$.

The simulation results of this extended model are presented in Figure~\ref{fig:bmu-model-2}. While the cell density profiles at the back of the \bmu\ have not changed qualitatively, the front of the \bmu\ now also exhibits structured profiles, clearly delineating a transition from the resorption zone (predominantly populated with \oca s and \ocm s) to the reversal zone (predominantly populated with precursor cells). This structure reproduces the histologically expected cell profiles of a cortical \bmu\ as schematically depicted in Figure~\ref{fig:schematic-bmu}.

The concentration profiles of \opg, \rankl\ and \tgfb\ are also plotted in Figure~\ref{fig:bmu-model-2}, with their heights scaled to correspond to the maximum of the \oba, \obp\ and \oca\ density profiles, respectively. One can see that \tgfb\ has slightly diffused to the back of \oca s and that its decline towards the back coincides with the onset of \oba s. As a result, there is no overlap of \oba s with \oca s. Since transport of soluble \opg\ has been assumed slow compared to its reaction processes (see Eq.~\eqref{fast-binding}), \opg\ is found mainly in the same location as \oba s that express it. On the other hand, while \rankl\ is bound to the membrane of \obp s, the \rankl\ concentration profile is clearly shifted towards the front of the \obp\ density profile, overlapping in particular with \ocp\ and \ocm\ cells. This is due to \opg\ inhibiting most of the available \rankl\ ligands at the back of the \obp\ profile.
With this observation, the role of \opg\ takes on a new fundamental meaning.


\section{Concluding Remarks}\label{sec:conclusions}
In this paper, we have developed a spatio-temporal mathematical model of cortical \bmu\ remodelling based on fundamental material balance equations expressed for different bone cell types. This model is an extension of the purely temporal model of Pivonka \etal\ \cite{pivonka-etal1}. \Tgfb, the \rank--\rankl--\opg\ pathway and \pth\ are explicitly taken into account in the model, and mass action kinetics is used to describe the reaction rates between ligands and their receptors. The resulting system of (nonlinear) PDEs is solved for the cell densities and regulatory factor concentrations in one dimension, corresponding to the longitudinal axis of a cortical \bmu\ (see Figure~\ref{fig:schematic-bmu}).

We find that the cell interaction pathways in the model are able to explain the emergence of a multi-cellular travelling wave that develops structured profiles moving without changing shape. The spatial structure of these steady profiles corresponds to the known organisation of bone cells in a typical cortical \bmu. It clearly delineates a resorption zone at the front, followed by a reversal zone, and a formation zone at the back. Several properties of the cell density profiles have been linked theoretically to kinetic properties of the cells in the model, such as differentiation and elimination rates. These rates may thus be directly inferred from the measurement of cell counts in serial histological sections taken at a particular time point.

It is experimentally known that several maturation stages of osteoclasts and osteoblasts have different expression patterns and responses in the \tgfb\ and \rank--\rankl--\opg\ pathways. Our model shows that this heterogeneity is essential to ensure a functional \bmu-remodelling process with segregated but coordinated zones of resorption and formation, in particular:
\begin{itemize}
\item \Tgfb\ plays a central role in modulating cell responses as soon as bone is resorbed. It moderates osteoclastic resorption and initiates osteoblastic commitment of mesenchymal cells once it has diffused from the resorption zone to the reversal zone. Furthermore, high concentration of \tgfb\ towards the front of the \bmu\ helps prevent osteoblasts from becoming activated prematurely, or near active osteoclasts.

\item The fact that \rankl\ is bound to the membrane of pre-osteoblasts helps ensure that osteoclasts do not initiate resorption without the presence of bone-refilling precursor cells. By shielding the availability of free \rankl\ on \obp s, the expression of soluble \opg\ by active osteoblasts in the formation zone provides a mechanism to ``shift'' the peak of the concentration profile of free \rankl\ towards the front of the \bmu, ahead of the peak of the \obp\ population that expresses \rankl, and thus prevents activation of osteoclasts where new bone is being laid down. Our model shows that changing the \rankl\ and \opg\ expression pattern fails to coordinate \oca\ formation properly.

\item The various motile properties of bone cells are absolutely crucial for the spatial organisation of the cells into a cortical \bmu, both in the formation zone and in the resorption zone. In particular, our results suggest that osteoclasts migrate forward at various rates depending on their maturation, and corroborate the observation that osteoblasts are quasi-stationary with respect to bone. The importance of bone cell motile properties is expected to play a critical role in the balance between bone resorption and bone formation, both in health and disease.
\end{itemize}

While our one-dimensional model is capable of reproducing important features of cortical \bmu s, there are a variety of possible improvements that could be addressed using the general formulation of the model presented in this paper. Solving the system of PDEs in higher spatial dimensions could be used to study how cells distribute themselves in transverse cross-sections of the \bmu. Other specific cell interaction pathways could be included as needed and studied in detail on the basis of the present model. We note that investigating initiation and termination signalling mechanisms of cortical \bmu s is very important to fully understand bone homeostasis during remodelling, and will be the subject of future work.

\subsubsection*{Acknowledgments}
This work is supported by the Australian Research Council grant ARC-DP-0879466 (``Bone regulation---cell interactions to disease'').

\end{document}

%% file: profiles-modified-root-model-bone-frame-at-0d.tex
\begingroup
  \makeatletter
  \providecommand\color[2][]{%
    \GenericError{(gnuplot) \space\space\space\@spaces}{%
      Package color not loaded in conjunction with
      terminal option `colourtext'%
    }{See the gnuplot documentation for explanation.%
    }{Either use 'blacktext' in gnuplot or load the package
      color.sty in LaTeX.}%
    \renewcommand\color[2][]{}%
  }%
  \providecommand\includegraphics[2][]{%
    \GenericError{(gnuplot) \space\space\space\@spaces}{%
      Package graphicx or graphics not loaded%
    }{See the gnuplot documentation for explanation.%
    }{The gnuplot epslatex terminal needs graphicx.sty or graphics.sty.}%
    \renewcommand\includegraphics[2][]{}%
  }%
  \providecommand\rotatebox[2]{#2}%
  \@ifundefined{ifGPcolor}{%
    \newif\ifGPcolor
    \GPcolortrue
  }{}%
  \@ifundefined{ifGPblacktext}{%
    \newif\ifGPblacktext
    \GPblacktexttrue
  }{}%
  \let\gplgaddtomacro\g@addto@macro
  \gdef\gplbacktext{}%
  \gdef\gplfronttext{}%
  \makeatother
  \ifGPblacktext
    \def\colorrgb#1{}%
    \def\colorgray#1{}%
  \else
    \ifGPcolor
      \def\colorrgb#1{\color[rgb]{#1}}%
      \def\colorgray#1{\color[gray]{#1}}%
      \expandafter\def\csname LTw\endcsname{\color{white}}%
      \expandafter\def\csname LTb\endcsname{\color{black}}%
      \expandafter\def\csname LTa\endcsname{\color{black}}%
      \expandafter\def\csname LT0\endcsname{\color[rgb]{1,0,0}}%
      \expandafter\def\csname LT1\endcsname{\color[rgb]{0,1,0}}%
      \expandafter\def\csname LT2\endcsname{\color[rgb]{0,0,1}}%
      \expandafter\def\csname LT3\endcsname{\color[rgb]{1,0,1}}%
      \expandafter\def\csname LT4\endcsname{\color[rgb]{0,1,1}}%
      \expandafter\def\csname LT5\endcsname{\color[rgb]{1,1,0}}%
      \expandafter\def\csname LT6\endcsname{\color[rgb]{0,0,0}}%
      \expandafter\def\csname LT7\endcsname{\color[rgb]{1,0.3,0}}%
      \expandafter\def\csname LT8\endcsname{\color[rgb]{0.5,0.5,0.5}}%
    \else
      \def\colorrgb#1{\color{black}}%
      \def\colorgray#1{\color[gray]{#1}}%
      \expandafter\def\csname LTw\endcsname{\color{white}}%
      \expandafter\def\csname LTb\endcsname{\color{black}}%
      \expandafter\def\csname LTa\endcsname{\color{black}}%
      \expandafter\def\csname LT0\endcsname{\color{black}}%
      \expandafter\def\csname LT1\endcsname{\color{black}}%
      \expandafter\def\csname LT2\endcsname{\color{black}}%
      \expandafter\def\csname LT3\endcsname{\color{black}}%
      \expandafter\def\csname LT4\endcsname{\color{black}}%
      \expandafter\def\csname LT5\endcsname{\color{black}}%
      \expandafter\def\csname LT6\endcsname{\color{black}}%
      \expandafter\def\csname LT7\endcsname{\color{black}}%
      \expandafter\def\csname LT8\endcsname{\color{black}}%
    \fi
  \fi
  \setlength{\unitlength}{0.0500bp}%
  \begin{picture}(4438.00,2160.00)%
    \gplgaddtomacro\gplbacktext{%
      \csname LTb\endcsname%
      \put(791,15){\makebox(0,0)[r]{\strut{}\footnotesize $0$}}%
      \put(791,365){\makebox(0,0)[r]{\strut{}\footnotesize $0.0005$}}%
      \put(791,715){\makebox(0,0)[r]{\strut{}\footnotesize $0.001$}}%
      \put(791,1066){\makebox(0,0)[r]{\strut{}\footnotesize $0.0015$}}%
      \put(791,1416){\makebox(0,0)[r]{\strut{}\footnotesize $0.002$}}%
      \put(791,1766){\makebox(0,0)[r]{\strut{}\footnotesize $0.0025$}}%
      \put(911,-185){\makebox(0,0){\strut{}}}%
      \put(1612,-185){\makebox(0,0){\strut{}}}%
      \put(2312,-185){\makebox(0,0){\strut{}}}%
      \put(3013,-185){\makebox(0,0){\strut{}}}%
      \put(3713,-185){\makebox(0,0){\strut{}}}%
      \put(4414,-185){\makebox(0,0){\strut{}}}%
      \put(91,890){\rotatebox{90}{\makebox(0,0){\strut{}\small Cell densities [pM]}}}%
      \put(2662,-285){\makebox(0,0){\strut{}}}%
      \put(2662,2066){\makebox(0,0){\strut{}\small Evolution of the cell profiles (bone frame)}}%
      \put(4361,1626){\makebox(0,0)[r]{\strut{}\footnotesize$t=0$ days}}%
    }%
    \gplgaddtomacro\gplfronttext{%
      \csname LTb\endcsname%
      \put(1764,1578){\makebox(0,0)[l]{\strut{}\footnotesize\oba}}%
      \csname LTb\endcsname%
      \put(1764,1378){\makebox(0,0)[l]{\strut{}\footnotesize\obp}}%
      \csname LTb\endcsname%
      \put(1764,1178){\makebox(0,0)[l]{\strut{}\footnotesize\oca}}%
      \csname LTb\endcsname%
      \put(1764,978){\makebox(0,0)[l]{\strut{}\footnotesize\obu\ and \ocp}}%
    }%
    \gplbacktext
    \put(0,0){\includegraphics{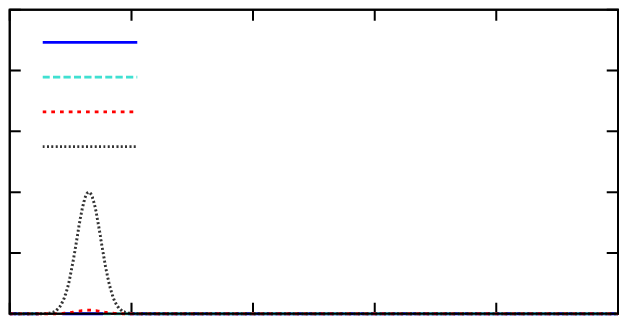}}%
    \gplfronttext
  \end{picture}%
\endgroup

%% file: profiles-modified-root-model-bone-frame-at-3d.tex
\begingroup
  \makeatletter
  \providecommand\color[2][]{%
    \GenericError{(gnuplot) \space\space\space\@spaces}{%
      Package color not loaded in conjunction with
      terminal option `colourtext'%
    }{See the gnuplot documentation for explanation.%
    }{Either use 'blacktext' in gnuplot or load the package
      color.sty in LaTeX.}%
    \renewcommand\color[2][]{}%
  }%
  \providecommand\includegraphics[2][]{%
    \GenericError{(gnuplot) \space\space\space\@spaces}{%
      Package graphicx or graphics not loaded%
    }{See the gnuplot documentation for explanation.%
    }{The gnuplot epslatex terminal needs graphicx.sty or graphics.sty.}%
    \renewcommand\includegraphics[2][]{}%
  }%
  \providecommand\rotatebox[2]{#2}%
  \@ifundefined{ifGPcolor}{%
    \newif\ifGPcolor
    \GPcolortrue
  }{}%
  \@ifundefined{ifGPblacktext}{%
    \newif\ifGPblacktext
    \GPblacktexttrue
  }{}%
  \let\gplgaddtomacro\g@addto@macro
  \gdef\gplbacktext{}%
  \gdef\gplfronttext{}%
  \makeatother
  \ifGPblacktext
    \def\colorrgb#1{}%
    \def\colorgray#1{}%
  \else
    \ifGPcolor
      \def\colorrgb#1{\color[rgb]{#1}}%
      \def\colorgray#1{\color[gray]{#1}}%
      \expandafter\def\csname LTw\endcsname{\color{white}}%
      \expandafter\def\csname LTb\endcsname{\color{black}}%
      \expandafter\def\csname LTa\endcsname{\color{black}}%
      \expandafter\def\csname LT0\endcsname{\color[rgb]{1,0,0}}%
      \expandafter\def\csname LT1\endcsname{\color[rgb]{0,1,0}}%
      \expandafter\def\csname LT2\endcsname{\color[rgb]{0,0,1}}%
      \expandafter\def\csname LT3\endcsname{\color[rgb]{1,0,1}}%
      \expandafter\def\csname LT4\endcsname{\color[rgb]{0,1,1}}%
      \expandafter\def\csname LT5\endcsname{\color[rgb]{1,1,0}}%
      \expandafter\def\csname LT6\endcsname{\color[rgb]{0,0,0}}%
      \expandafter\def\csname LT7\endcsname{\color[rgb]{1,0.3,0}}%
      \expandafter\def\csname LT8\endcsname{\color[rgb]{0.5,0.5,0.5}}%
    \else
      \def\colorrgb#1{\color{black}}%
      \def\colorgray#1{\color[gray]{#1}}%
      \expandafter\def\csname LTw\endcsname{\color{white}}%
      \expandafter\def\csname LTb\endcsname{\color{black}}%
      \expandafter\def\csname LTa\endcsname{\color{black}}%
      \expandafter\def\csname LT0\endcsname{\color{black}}%
      \expandafter\def\csname LT1\endcsname{\color{black}}%
      \expandafter\def\csname LT2\endcsname{\color{black}}%
      \expandafter\def\csname LT3\endcsname{\color{black}}%
      \expandafter\def\csname LT4\endcsname{\color{black}}%
      \expandafter\def\csname LT5\endcsname{\color{black}}%
      \expandafter\def\csname LT6\endcsname{\color{black}}%
      \expandafter\def\csname LT7\endcsname{\color{black}}%
      \expandafter\def\csname LT8\endcsname{\color{black}}%
    \fi
  \fi
  \setlength{\unitlength}{0.0500bp}%
  \begin{picture}(4438.00,1828.00)%
    \gplgaddtomacro\gplbacktext{%
      \csname LTb\endcsname%
      \put(791,38){\makebox(0,0)[r]{\strut{}\footnotesize $0$}}%
      \put(791,388){\makebox(0,0)[r]{\strut{}\footnotesize $0.0005$}}%
      \put(791,738){\makebox(0,0)[r]{\strut{}\footnotesize $0.001$}}%
      \put(791,1089){\makebox(0,0)[r]{\strut{}\footnotesize $0.0015$}}%
      \put(791,1439){\makebox(0,0)[r]{\strut{}\footnotesize $0.002$}}%
      \put(791,1789){\makebox(0,0)[r]{\strut{}\footnotesize $0.0025$}}%
      \put(911,-162){\makebox(0,0){\strut{}}}%
      \put(1612,-162){\makebox(0,0){\strut{}}}%
      \put(2312,-162){\makebox(0,0){\strut{}}}%
      \put(3013,-162){\makebox(0,0){\strut{}}}%
      \put(3713,-162){\makebox(0,0){\strut{}}}%
      \put(4414,-162){\makebox(0,0){\strut{}}}%
      \put(91,913){\rotatebox{90}{\makebox(0,0){\strut{}\small Cell densities [pM]}}}%
      \put(2662,-262){\makebox(0,0){\strut{}}}%
      \put(4361,1649){\makebox(0,0)[r]{\strut{}\footnotesize$t=3$ days}}%
    }%
    \gplgaddtomacro\gplfronttext{%
    }%
    \gplbacktext
    \put(0,0){\includegraphics{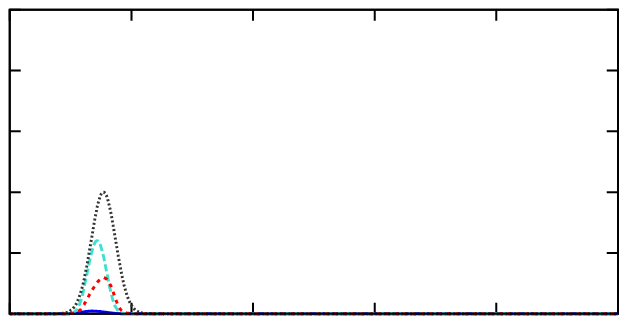}}%
    \gplfronttext
  \end{picture}%
\endgroup

%% file: profiles-modified-root-model-bone-frame-at-20d.tex
\begingroup
  \makeatletter
  \providecommand\color[2][]{%
    \GenericError{(gnuplot) \space\space\space\@spaces}{%
      Package color not loaded in conjunction with
      terminal option `colourtext'%
    }{See the gnuplot documentation for explanation.%
    }{Either use 'blacktext' in gnuplot or load the package
      color.sty in LaTeX.}%
    \renewcommand\color[2][]{}%
  }%
  \providecommand\includegraphics[2][]{%
    \GenericError{(gnuplot) \space\space\space\@spaces}{%
      Package graphicx or graphics not loaded%
    }{See the gnuplot documentation for explanation.%
    }{The gnuplot epslatex terminal needs graphicx.sty or graphics.sty.}%
    \renewcommand\includegraphics[2][]{}%
  }%
  \providecommand\rotatebox[2]{#2}%
  \@ifundefined{ifGPcolor}{%
    \newif\ifGPcolor
    \GPcolortrue
  }{}%
  \@ifundefined{ifGPblacktext}{%
    \newif\ifGPblacktext
    \GPblacktexttrue
  }{}%
  \let\gplgaddtomacro\g@addto@macro
  \gdef\gplbacktext{}%
  \gdef\gplfronttext{}%
  \makeatother
  \ifGPblacktext
    \def\colorrgb#1{}%
    \def\colorgray#1{}%
  \else
    \ifGPcolor
      \def\colorrgb#1{\color[rgb]{#1}}%
      \def\colorgray#1{\color[gray]{#1}}%
      \expandafter\def\csname LTw\endcsname{\color{white}}%
      \expandafter\def\csname LTb\endcsname{\color{black}}%
      \expandafter\def\csname LTa\endcsname{\color{black}}%
      \expandafter\def\csname LT0\endcsname{\color[rgb]{1,0,0}}%
      \expandafter\def\csname LT1\endcsname{\color[rgb]{0,1,0}}%
      \expandafter\def\csname LT2\endcsname{\color[rgb]{0,0,1}}%
      \expandafter\def\csname LT3\endcsname{\color[rgb]{1,0,1}}%
      \expandafter\def\csname LT4\endcsname{\color[rgb]{0,1,1}}%
      \expandafter\def\csname LT5\endcsname{\color[rgb]{1,1,0}}%
      \expandafter\def\csname LT6\endcsname{\color[rgb]{0,0,0}}%
      \expandafter\def\csname LT7\endcsname{\color[rgb]{1,0.3,0}}%
      \expandafter\def\csname LT8\endcsname{\color[rgb]{0.5,0.5,0.5}}%
    \else
      \def\colorrgb#1{\color{black}}%
      \def\colorgray#1{\color[gray]{#1}}%
      \expandafter\def\csname LTw\endcsname{\color{white}}%
      \expandafter\def\csname LTb\endcsname{\color{black}}%
      \expandafter\def\csname LTa\endcsname{\color{black}}%
      \expandafter\def\csname LT0\endcsname{\color{black}}%
      \expandafter\def\csname LT1\endcsname{\color{black}}%
      \expandafter\def\csname LT2\endcsname{\color{black}}%
      \expandafter\def\csname LT3\endcsname{\color{black}}%
      \expandafter\def\csname LT4\endcsname{\color{black}}%
      \expandafter\def\csname LT5\endcsname{\color{black}}%
      \expandafter\def\csname LT6\endcsname{\color{black}}%
      \expandafter\def\csname LT7\endcsname{\color{black}}%
      \expandafter\def\csname LT8\endcsname{\color{black}}%
    \fi
  \fi
  \setlength{\unitlength}{0.0500bp}%
  \begin{picture}(4438.00,1828.00)%
    \gplgaddtomacro\gplbacktext{%
      \csname LTb\endcsname%
      \put(791,38){\makebox(0,0)[r]{\strut{}\footnotesize $0$}}%
      \put(791,388){\makebox(0,0)[r]{\strut{}\footnotesize $0.0005$}}%
      \put(791,738){\makebox(0,0)[r]{\strut{}\footnotesize $0.001$}}%
      \put(791,1089){\makebox(0,0)[r]{\strut{}\footnotesize $0.0015$}}%
      \put(791,1439){\makebox(0,0)[r]{\strut{}\footnotesize $0.002$}}%
      \put(791,1789){\makebox(0,0)[r]{\strut{}\footnotesize $0.0025$}}%
      \put(911,-162){\makebox(0,0){\strut{}}}%
      \put(1612,-162){\makebox(0,0){\strut{}}}%
      \put(2312,-162){\makebox(0,0){\strut{}}}%
      \put(3013,-162){\makebox(0,0){\strut{}}}%
      \put(3713,-162){\makebox(0,0){\strut{}}}%
      \put(4414,-162){\makebox(0,0){\strut{}}}%
      \put(91,913){\rotatebox{90}{\makebox(0,0){\strut{}\small Cell densities [pM]}}}%
      \put(2662,-262){\makebox(0,0){\strut{}}}%
      \put(4361,1649){\makebox(0,0)[r]{\strut{}\footnotesize$t=20$ days}}%
    }%
    \gplgaddtomacro\gplfronttext{%
    }%
    \gplbacktext
    \put(0,0){\includegraphics{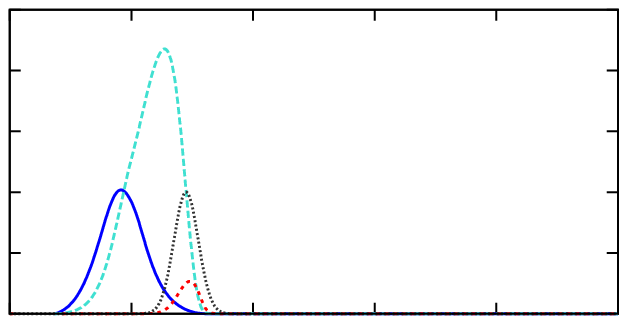}}%
    \gplfronttext
  \end{picture}%
\endgroup

%% file: profiles-modified-root-model-bone-frame-at-100d-with-inset.tex
\begingroup
  \makeatletter
  \providecommand\color[2][]{%
    \GenericError{(gnuplot) \space\space\space\@spaces}{%
      Package color not loaded in conjunction with
      terminal option `colourtext'%
    }{See the gnuplot documentation for explanation.%
    }{Either use 'blacktext' in gnuplot or load the package
      color.sty in LaTeX.}%
    \renewcommand\color[2][]{}%
  }%
  \providecommand\includegraphics[2][]{%
    \GenericError{(gnuplot) \space\space\space\@spaces}{%
      Package graphicx or graphics not loaded%
    }{See the gnuplot documentation for explanation.%
    }{The gnuplot epslatex terminal needs graphicx.sty or graphics.sty.}%
    \renewcommand\includegraphics[2][]{}%
  }%
  \providecommand\rotatebox[2]{#2}%
  \@ifundefined{ifGPcolor}{%
    \newif\ifGPcolor
    \GPcolortrue
  }{}%
  \@ifundefined{ifGPblacktext}{%
    \newif\ifGPblacktext
    \GPblacktexttrue
  }{}%
  \let\gplgaddtomacro\g@addto@macro
  \gdef\gplbacktext{}%
  \gdef\gplfronttext{}%
  \makeatother
  \ifGPblacktext
    \def\colorrgb#1{}%
    \def\colorgray#1{}%
  \else
    \ifGPcolor
      \def\colorrgb#1{\color[rgb]{#1}}%
      \def\colorgray#1{\color[gray]{#1}}%
      \expandafter\def\csname LTw\endcsname{\color{white}}%
      \expandafter\def\csname LTb\endcsname{\color{black}}%
      \expandafter\def\csname LTa\endcsname{\color{black}}%
      \expandafter\def\csname LT0\endcsname{\color[rgb]{1,0,0}}%
      \expandafter\def\csname LT1\endcsname{\color[rgb]{0,1,0}}%
      \expandafter\def\csname LT2\endcsname{\color[rgb]{0,0,1}}%
      \expandafter\def\csname LT3\endcsname{\color[rgb]{1,0,1}}%
      \expandafter\def\csname LT4\endcsname{\color[rgb]{0,1,1}}%
      \expandafter\def\csname LT5\endcsname{\color[rgb]{1,1,0}}%
      \expandafter\def\csname LT6\endcsname{\color[rgb]{0,0,0}}%
      \expandafter\def\csname LT7\endcsname{\color[rgb]{1,0.3,0}}%
      \expandafter\def\csname LT8\endcsname{\color[rgb]{0.5,0.5,0.5}}%
    \else
      \def\colorrgb#1{\color{black}}%
      \def\colorgray#1{\color[gray]{#1}}%
      \expandafter\def\csname LTw\endcsname{\color{white}}%
      \expandafter\def\csname LTb\endcsname{\color{black}}%
      \expandafter\def\csname LTa\endcsname{\color{black}}%
      \expandafter\def\csname LT0\endcsname{\color{black}}%
      \expandafter\def\csname LT1\endcsname{\color{black}}%
      \expandafter\def\csname LT2\endcsname{\color{black}}%
      \expandafter\def\csname LT3\endcsname{\color{black}}%
      \expandafter\def\csname LT4\endcsname{\color{black}}%
      \expandafter\def\csname LT5\endcsname{\color{black}}%
      \expandafter\def\csname LT6\endcsname{\color{black}}%
      \expandafter\def\csname LT7\endcsname{\color{black}}%
      \expandafter\def\csname LT8\endcsname{\color{black}}%
    \fi
  \fi
  \setlength{\unitlength}{0.0500bp}%
  \begin{picture}(4438.00,1828.00)%
    \gplgaddtomacro\gplbacktext{%
      \csname LTb\endcsname%
      \put(791,38){\makebox(0,0)[r]{\strut{}\footnotesize $0$}}%
      \put(791,388){\makebox(0,0)[r]{\strut{}\footnotesize $0.0005$}}%
      \put(791,738){\makebox(0,0)[r]{\strut{}\footnotesize $0.001$}}%
      \put(791,1089){\makebox(0,0)[r]{\strut{}\footnotesize $0.0015$}}%
      \put(791,1439){\makebox(0,0)[r]{\strut{}\footnotesize $0.002$}}%
      \put(791,1789){\makebox(0,0)[r]{\strut{}\footnotesize $0.0025$}}%
      \put(911,-162){\makebox(0,0){\strut{}}}%
      \put(1612,-162){\makebox(0,0){\strut{}}}%
      \put(2312,-162){\makebox(0,0){\strut{}}}%
      \put(3013,-162){\makebox(0,0){\strut{}}}%
      \put(3713,-162){\makebox(0,0){\strut{}}}%
      \put(4414,-162){\makebox(0,0){\strut{}}}%
      \put(91,913){\rotatebox{90}{\makebox(0,0){\strut{}\small Cell densities [pM]}}}%
      \put(2662,-262){\makebox(0,0){\strut{}}}%
      \put(4361,1649){\makebox(0,0)[r]{\strut{}\footnotesize$t=100$ days}}%
    }%
    \gplgaddtomacro\gplfronttext{%
    }%
    \gplgaddtomacro\gplbacktext{%
      \csname LTb\endcsname%
      \put(1197,590){\rotatebox{-30}{\makebox(0,0)[l]{\strut{}\scriptsize $-4000$}}}%
      \put(1578,590){\rotatebox{-30}{\makebox(0,0)[l]{\strut{}\scriptsize $-3000$}}}%
      \put(1959,590){\rotatebox{-30}{\makebox(0,0)[l]{\strut{}\scriptsize $-2000$}}}%
      \put(2340,590){\rotatebox{-30}{\makebox(0,0)[l]{\strut{}\scriptsize $-1000$}}}%
      \put(2721,590){\rotatebox{-30}{\makebox(0,0)[l]{\strut{}\scriptsize $0$}}}%
      \put(2781,640){\makebox(0,0)[l]{\strut{}\scriptsize $10^{-9}$}}%
      \put(2781,798){\makebox(0,0)[l]{\strut{}\scriptsize $10^{-8}$}}%
      \put(2781,956){\makebox(0,0)[l]{\strut{}\scriptsize $10^{-7}$}}%
      \put(2781,1115){\makebox(0,0)[l]{\strut{}\scriptsize $10^{-6}$}}%
      \put(2781,1273){\makebox(0,0)[l]{\strut{}\scriptsize $10^{-5}$}}%
      \put(2781,1431){\makebox(0,0)[l]{\strut{}\scriptsize $10^{-4}$}}%
      \put(2781,1589){\makebox(0,0)[l]{\strut{}\scriptsize $10^{-3}$}}%
      \put(3166,1170){\rotatebox{90}{\makebox(0,0){\strut{}}}}%
      \put(1863,-60){\makebox(0,0){\strut{}}}%
    }%
    \gplgaddtomacro\gplfronttext{%
    }%
    \gplbacktext
    \put(0,0){\includegraphics{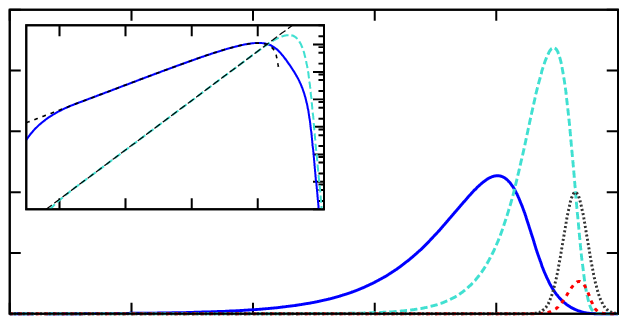}}%
    \gplfronttext
  \end{picture}%
\endgroup

%% file: bmu-cavity-sketch-bone-frame-at-100d.tex
\begingroup
  \makeatletter
  \providecommand\color[2][]{%
    \GenericError{(gnuplot) \space\space\space\@spaces}{%
      Package color not loaded in conjunction with
      terminal option `colourtext'%
    }{See the gnuplot documentation for explanation.%
    }{Either use 'blacktext' in gnuplot or load the package
      color.sty in LaTeX.}%
    \renewcommand\color[2][]{}%
  }%
  \providecommand\includegraphics[2][]{%
    \GenericError{(gnuplot) \space\space\space\@spaces}{%
      Package graphicx or graphics not loaded%
    }{See the gnuplot documentation for explanation.%
    }{The gnuplot epslatex terminal needs graphicx.sty or graphics.sty.}%
    \renewcommand\includegraphics[2][]{}%
  }%
  \providecommand\rotatebox[2]{#2}%
  \@ifundefined{ifGPcolor}{%
    \newif\ifGPcolor
    \GPcolortrue
  }{}%
  \@ifundefined{ifGPblacktext}{%
    \newif\ifGPblacktext
    \GPblacktexttrue
  }{}%
  \let\gplgaddtomacro\g@addto@macro
  \gdef\gplbacktext{}%
  \gdef\gplfronttext{}%
  \makeatother
  \ifGPblacktext
    \def\colorrgb#1{}%
    \def\colorgray#1{}%
  \else
    \ifGPcolor
      \def\colorrgb#1{\color[rgb]{#1}}%
      \def\colorgray#1{\color[gray]{#1}}%
      \expandafter\def\csname LTw\endcsname{\color{white}}%
      \expandafter\def\csname LTb\endcsname{\color{black}}%
      \expandafter\def\csname LTa\endcsname{\color{black}}%
      \expandafter\def\csname LT0\endcsname{\color[rgb]{1,0,0}}%
      \expandafter\def\csname LT1\endcsname{\color[rgb]{0,1,0}}%
      \expandafter\def\csname LT2\endcsname{\color[rgb]{0,0,1}}%
      \expandafter\def\csname LT3\endcsname{\color[rgb]{1,0,1}}%
      \expandafter\def\csname LT4\endcsname{\color[rgb]{0,1,1}}%
      \expandafter\def\csname LT5\endcsname{\color[rgb]{1,1,0}}%
      \expandafter\def\csname LT6\endcsname{\color[rgb]{0,0,0}}%
      \expandafter\def\csname LT7\endcsname{\color[rgb]{1,0.3,0}}%
      \expandafter\def\csname LT8\endcsname{\color[rgb]{0.5,0.5,0.5}}%
    \else
      \def\colorrgb#1{\color{black}}%
      \def\colorgray#1{\color[gray]{#1}}%
      \expandafter\def\csname LTw\endcsname{\color{white}}%
      \expandafter\def\csname LTb\endcsname{\color{black}}%
      \expandafter\def\csname LTa\endcsname{\color{black}}%
      \expandafter\def\csname LT0\endcsname{\color{black}}%
      \expandafter\def\csname LT1\endcsname{\color{black}}%
      \expandafter\def\csname LT2\endcsname{\color{black}}%
      \expandafter\def\csname LT3\endcsname{\color{black}}%
      \expandafter\def\csname LT4\endcsname{\color{black}}%
      \expandafter\def\csname LT5\endcsname{\color{black}}%
      \expandafter\def\csname LT6\endcsname{\color{black}}%
      \expandafter\def\csname LT7\endcsname{\color{black}}%
      \expandafter\def\csname LT8\endcsname{\color{black}}%
    \fi
  \fi
  \setlength{\unitlength}{0.0500bp}%
  \begin{picture}(4438.00,886.00)%
    \gplgaddtomacro\gplbacktext{%
      \csname LTb\endcsname%
      \put(2663,52){\makebox(0,0){\strut{}\small $x$ axis [$\um$]}}%
    }%
    \gplgaddtomacro\gplfronttext{%
      \csname LTb\endcsname%
      \put(792,602){\makebox(0,0)[r]{\strut{}\footnotesize $-100$}}%
      \csname LTb\endcsname%
      \put(792,703){\makebox(0,0)[r]{\strut{}\footnotesize $0$}}%
      \csname LTb\endcsname%
      \put(792,803){\makebox(0,0)[r]{\strut{}\footnotesize $100$}}%
      \csname LTb\endcsname%
      \put(912,352){\makebox(0,0){\strut{}\footnotesize $-5000$}}%
      \csname LTb\endcsname%
      \put(1612,352){\makebox(0,0){\strut{}\footnotesize $-4000$}}%
      \csname LTb\endcsname%
      \put(2313,352){\makebox(0,0){\strut{}\footnotesize $-3000$}}%
      \csname LTb\endcsname%
      \put(3013,352){\makebox(0,0){\strut{}\footnotesize $-2000$}}%
      \csname LTb\endcsname%
      \put(3714,352){\makebox(0,0){\strut{}\footnotesize $-1000$}}%
      \csname LTb\endcsname%
      \put(4414,352){\makebox(0,0){\strut{}\footnotesize $0$}}%
    }%
    \gplbacktext
    \put(0,0){\includegraphics{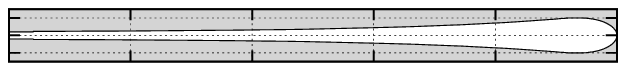}}%
    \gplfronttext
  \end{picture}%
\endgroup

%% file: profiles-modified-root-model-allFluxesnv-at-100d.tex
\begingroup
  \makeatletter
  \providecommand\color[2][]{%
    \GenericError{(gnuplot) \space\space\space\@spaces}{%
      Package color not loaded in conjunction with
      terminal option `colourtext'%
    }{See the gnuplot documentation for explanation.%
    }{Either use 'blacktext' in gnuplot or load the package
      color.sty in LaTeX.}%
    \renewcommand\color[2][]{}%
  }%
  \providecommand\includegraphics[2][]{%
    \GenericError{(gnuplot) \space\space\space\@spaces}{%
      Package graphicx or graphics not loaded%
    }{See the gnuplot documentation for explanation.%
    }{The gnuplot epslatex terminal needs graphicx.sty or graphics.sty.}%
    \renewcommand\includegraphics[2][]{}%
  }%
  \providecommand\rotatebox[2]{#2}%
  \@ifundefined{ifGPcolor}{%
    \newif\ifGPcolor
    \GPcolortrue
  }{}%
  \@ifundefined{ifGPblacktext}{%
    \newif\ifGPblacktext
    \GPblacktexttrue
  }{}%
  \let\gplgaddtomacro\g@addto@macro
  \gdef\gplbacktext{}%
  \gdef\gplfronttext{}%
  \makeatother
  \ifGPblacktext
    \def\colorrgb#1{}%
    \def\colorgray#1{}%
  \else
    \ifGPcolor
      \def\colorrgb#1{\color[rgb]{#1}}%
      \def\colorgray#1{\color[gray]{#1}}%
      \expandafter\def\csname LTw\endcsname{\color{white}}%
      \expandafter\def\csname LTb\endcsname{\color{black}}%
      \expandafter\def\csname LTa\endcsname{\color{black}}%
      \expandafter\def\csname LT0\endcsname{\color[rgb]{1,0,0}}%
      \expandafter\def\csname LT1\endcsname{\color[rgb]{0,1,0}}%
      \expandafter\def\csname LT2\endcsname{\color[rgb]{0,0,1}}%
      \expandafter\def\csname LT3\endcsname{\color[rgb]{1,0,1}}%
      \expandafter\def\csname LT4\endcsname{\color[rgb]{0,1,1}}%
      \expandafter\def\csname LT5\endcsname{\color[rgb]{1,1,0}}%
      \expandafter\def\csname LT6\endcsname{\color[rgb]{0,0,0}}%
      \expandafter\def\csname LT7\endcsname{\color[rgb]{1,0.3,0}}%
      \expandafter\def\csname LT8\endcsname{\color[rgb]{0.5,0.5,0.5}}%
    \else
      \def\colorrgb#1{\color{black}}%
      \def\colorgray#1{\color[gray]{#1}}%
      \expandafter\def\csname LTw\endcsname{\color{white}}%
      \expandafter\def\csname LTb\endcsname{\color{black}}%
      \expandafter\def\csname LTa\endcsname{\color{black}}%
      \expandafter\def\csname LT0\endcsname{\color{black}}%
      \expandafter\def\csname LT1\endcsname{\color{black}}%
      \expandafter\def\csname LT2\endcsname{\color{black}}%
      \expandafter\def\csname LT3\endcsname{\color{black}}%
      \expandafter\def\csname LT4\endcsname{\color{black}}%
      \expandafter\def\csname LT5\endcsname{\color{black}}%
      \expandafter\def\csname LT6\endcsname{\color{black}}%
      \expandafter\def\csname LT7\endcsname{\color{black}}%
      \expandafter\def\csname LT8\endcsname{\color{black}}%
    \fi
  \fi
  \setlength{\unitlength}{0.0500bp}%
  \begin{picture}(4438.00,2160.00)%
    \gplgaddtomacro\gplbacktext{%
      \csname LTb\endcsname%
      \put(791,15){\makebox(0,0)[r]{\strut{}\footnotesize $0$}}%
      \put(791,365){\makebox(0,0)[r]{\strut{}\footnotesize $0.0005$}}%
      \put(791,715){\makebox(0,0)[r]{\strut{}\footnotesize $0.001$}}%
      \put(791,1066){\makebox(0,0)[r]{\strut{}\footnotesize $0.0015$}}%
      \put(791,1416){\makebox(0,0)[r]{\strut{}\footnotesize $0.002$}}%
      \put(791,1766){\makebox(0,0)[r]{\strut{}\footnotesize $0.0025$}}%
      \put(911,-185){\makebox(0,0){\strut{}}}%
      \put(1612,-185){\makebox(0,0){\strut{}}}%
      \put(2312,-185){\makebox(0,0){\strut{}}}%
      \put(3013,-185){\makebox(0,0){\strut{}}}%
      \put(3713,-185){\makebox(0,0){\strut{}}}%
      \put(4414,-185){\makebox(0,0){\strut{}}}%
      \put(91,890){\rotatebox{90}{\makebox(0,0){\strut{}\small Cell densities [pM]}}}%
      \put(2662,-285){\makebox(0,0){\strut{}}}%
      \put(2662,2066){\makebox(0,0){\strut{}\small Cell profiles with wrong cellular fluxes}}%
    }%
    \gplgaddtomacro\gplfronttext{%
      \csname LTb\endcsname%
      \put(1764,1578){\makebox(0,0)[l]{\strut{}\footnotesize\oba}}%
      \csname LTb\endcsname%
      \put(1764,1378){\makebox(0,0)[l]{\strut{}\footnotesize\obp}}%
      \csname LTb\endcsname%
      \put(1764,1178){\makebox(0,0)[l]{\strut{}\footnotesize\oca}}%
      \csname LTb\endcsname%
      \put(1764,978){\makebox(0,0)[l]{\strut{}\footnotesize\obu\ and \ocp}}%
    }%
    \gplbacktext
    \put(0,0){\includegraphics{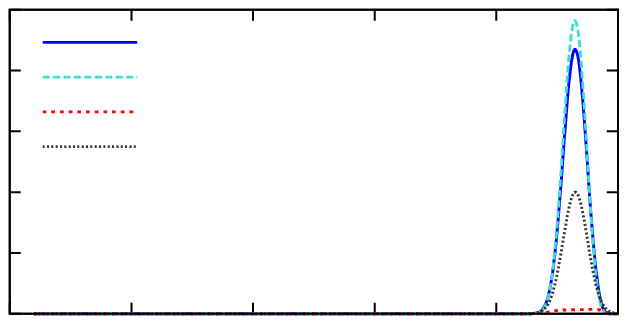}}%
    \gplfronttext
  \end{picture}%
\endgroup

%% file: profiles-modified-root-model-OpgRanklSwap-at-100d.tex
\begingroup
  \makeatletter
  \providecommand\color[2][]{%
    \GenericError{(gnuplot) \space\space\space\@spaces}{%
      Package color not loaded in conjunction with
      terminal option `colourtext'%
    }{See the gnuplot documentation for explanation.%
    }{Either use 'blacktext' in gnuplot or load the package
      color.sty in LaTeX.}%
    \renewcommand\color[2][]{}%
  }%
  \providecommand\includegraphics[2][]{%
    \GenericError{(gnuplot) \space\space\space\@spaces}{%
      Package graphicx or graphics not loaded%
    }{See the gnuplot documentation for explanation.%
    }{The gnuplot epslatex terminal needs graphicx.sty or graphics.sty.}%
    \renewcommand\includegraphics[2][]{}%
  }%
  \providecommand\rotatebox[2]{#2}%
  \@ifundefined{ifGPcolor}{%
    \newif\ifGPcolor
    \GPcolortrue
  }{}%
  \@ifundefined{ifGPblacktext}{%
    \newif\ifGPblacktext
    \GPblacktexttrue
  }{}%
  \let\gplgaddtomacro\g@addto@macro
  \gdef\gplbacktext{}%
  \gdef\gplfronttext{}%
  \makeatother
  \ifGPblacktext
    \def\colorrgb#1{}%
    \def\colorgray#1{}%
  \else
    \ifGPcolor
      \def\colorrgb#1{\color[rgb]{#1}}%
      \def\colorgray#1{\color[gray]{#1}}%
      \expandafter\def\csname LTw\endcsname{\color{white}}%
      \expandafter\def\csname LTb\endcsname{\color{black}}%
      \expandafter\def\csname LTa\endcsname{\color{black}}%
      \expandafter\def\csname LT0\endcsname{\color[rgb]{1,0,0}}%
      \expandafter\def\csname LT1\endcsname{\color[rgb]{0,1,0}}%
      \expandafter\def\csname LT2\endcsname{\color[rgb]{0,0,1}}%
      \expandafter\def\csname LT3\endcsname{\color[rgb]{1,0,1}}%
      \expandafter\def\csname LT4\endcsname{\color[rgb]{0,1,1}}%
      \expandafter\def\csname LT5\endcsname{\color[rgb]{1,1,0}}%
      \expandafter\def\csname LT6\endcsname{\color[rgb]{0,0,0}}%
      \expandafter\def\csname LT7\endcsname{\color[rgb]{1,0.3,0}}%
      \expandafter\def\csname LT8\endcsname{\color[rgb]{0.5,0.5,0.5}}%
    \else
      \def\colorrgb#1{\color{black}}%
      \def\colorgray#1{\color[gray]{#1}}%
      \expandafter\def\csname LTw\endcsname{\color{white}}%
      \expandafter\def\csname LTb\endcsname{\color{black}}%
      \expandafter\def\csname LTa\endcsname{\color{black}}%
      \expandafter\def\csname LT0\endcsname{\color{black}}%
      \expandafter\def\csname LT1\endcsname{\color{black}}%
      \expandafter\def\csname LT2\endcsname{\color{black}}%
      \expandafter\def\csname LT3\endcsname{\color{black}}%
      \expandafter\def\csname LT4\endcsname{\color{black}}%
      \expandafter\def\csname LT5\endcsname{\color{black}}%
      \expandafter\def\csname LT6\endcsname{\color{black}}%
      \expandafter\def\csname LT7\endcsname{\color{black}}%
      \expandafter\def\csname LT8\endcsname{\color{black}}%
    \fi
  \fi
  \setlength{\unitlength}{0.0500bp}%
  \begin{picture}(4438.00,2160.00)%
    \gplgaddtomacro\gplbacktext{%
      \csname LTb\endcsname%
      \put(791,15){\makebox(0,0)[r]{\strut{}\footnotesize $0$}}%
      \put(791,365){\makebox(0,0)[r]{\strut{}\footnotesize $0.0005$}}%
      \put(791,715){\makebox(0,0)[r]{\strut{}\footnotesize $0.001$}}%
      \put(791,1066){\makebox(0,0)[r]{\strut{}\footnotesize $0.0015$}}%
      \put(791,1416){\makebox(0,0)[r]{\strut{}\footnotesize $0.002$}}%
      \put(791,1766){\makebox(0,0)[r]{\strut{}\footnotesize $0.0025$}}%
      \put(911,-185){\makebox(0,0){\strut{}}}%
      \put(1612,-185){\makebox(0,0){\strut{}}}%
      \put(2312,-185){\makebox(0,0){\strut{}}}%
      \put(3013,-185){\makebox(0,0){\strut{}}}%
      \put(3713,-185){\makebox(0,0){\strut{}}}%
      \put(4414,-185){\makebox(0,0){\strut{}}}%
      \put(91,890){\rotatebox{90}{\makebox(0,0){\strut{}\small Cell densities [pM]}}}%
      \put(2662,-285){\makebox(0,0){\strut{}}}%
      \put(2422,2066){\makebox(0,0){\strut{}\small Cell profiles with \rankl\ and \opg\ expression swapped}}%
    }%
    \gplgaddtomacro\gplfronttext{%
      \csname LTb\endcsname%
      \put(1764,1578){\makebox(0,0)[l]{\strut{}\footnotesize\oba}}%
      \csname LTb\endcsname%
      \put(1764,1378){\makebox(0,0)[l]{\strut{}\footnotesize\obp}}%
      \csname LTb\endcsname%
      \put(1764,1178){\makebox(0,0)[l]{\strut{}\footnotesize\oca}}%
      \csname LTb\endcsname%
      \put(1764,978){\makebox(0,0)[l]{\strut{}\footnotesize\obu\ and \ocp}}%
    }%
    \gplbacktext
    \put(0,0){\includegraphics{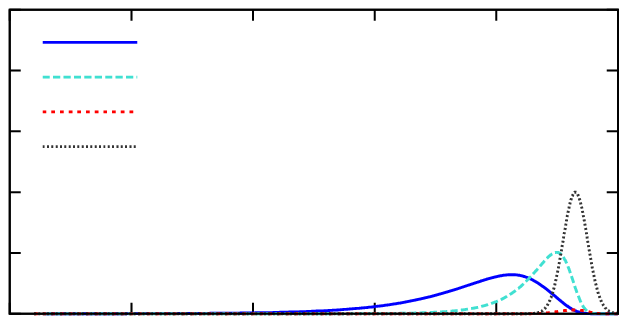}}%
    \gplfronttext
  \end{picture}%
\endgroup

%% file: profiles-model2-at-100d-cells.tex
\begingroup
  \makeatletter
  \providecommand\color[2][]{%
    \GenericError{(gnuplot) \space\space\space\@spaces}{%
      Package color not loaded in conjunction with
      terminal option `colourtext'%
    }{See the gnuplot documentation for explanation.%
    }{Either use 'blacktext' in gnuplot or load the package
      color.sty in LaTeX.}%
    \renewcommand\color[2][]{}%
  }%
  \providecommand\includegraphics[2][]{%
    \GenericError{(gnuplot) \space\space\space\@spaces}{%
      Package graphicx or graphics not loaded%
    }{See the gnuplot documentation for explanation.%
    }{The gnuplot epslatex terminal needs graphicx.sty or graphics.sty.}%
    \renewcommand\includegraphics[2][]{}%
  }%
  \providecommand\rotatebox[2]{#2}%
  \@ifundefined{ifGPcolor}{%
    \newif\ifGPcolor
    \GPcolortrue
  }{}%
  \@ifundefined{ifGPblacktext}{%
    \newif\ifGPblacktext
    \GPblacktexttrue
  }{}%
  \let\gplgaddtomacro\g@addto@macro
  \gdef\gplbacktext{}%
  \gdef\gplfronttext{}%
  \makeatother
  \ifGPblacktext
    \def\colorrgb#1{}%
    \def\colorgray#1{}%
  \else
    \ifGPcolor
      \def\colorrgb#1{\color[rgb]{#1}}%
      \def\colorgray#1{\color[gray]{#1}}%
      \expandafter\def\csname LTw\endcsname{\color{white}}%
      \expandafter\def\csname LTb\endcsname{\color{black}}%
      \expandafter\def\csname LTa\endcsname{\color{black}}%
      \expandafter\def\csname LT0\endcsname{\color[rgb]{1,0,0}}%
      \expandafter\def\csname LT1\endcsname{\color[rgb]{0,1,0}}%
      \expandafter\def\csname LT2\endcsname{\color[rgb]{0,0,1}}%
      \expandafter\def\csname LT3\endcsname{\color[rgb]{1,0,1}}%
      \expandafter\def\csname LT4\endcsname{\color[rgb]{0,1,1}}%
      \expandafter\def\csname LT5\endcsname{\color[rgb]{1,1,0}}%
      \expandafter\def\csname LT6\endcsname{\color[rgb]{0,0,0}}%
      \expandafter\def\csname LT7\endcsname{\color[rgb]{1,0.3,0}}%
      \expandafter\def\csname LT8\endcsname{\color[rgb]{0.5,0.5,0.5}}%
    \else
      \def\colorrgb#1{\color{black}}%
      \def\colorgray#1{\color[gray]{#1}}%
      \expandafter\def\csname LTw\endcsname{\color{white}}%
      \expandafter\def\csname LTb\endcsname{\color{black}}%
      \expandafter\def\csname LTa\endcsname{\color{black}}%
      \expandafter\def\csname LT0\endcsname{\color{black}}%
      \expandafter\def\csname LT1\endcsname{\color{black}}%
      \expandafter\def\csname LT2\endcsname{\color{black}}%
      \expandafter\def\csname LT3\endcsname{\color{black}}%
      \expandafter\def\csname LT4\endcsname{\color{black}}%
      \expandafter\def\csname LT5\endcsname{\color{black}}%
      \expandafter\def\csname LT6\endcsname{\color{black}}%
      \expandafter\def\csname LT7\endcsname{\color{black}}%
      \expandafter\def\csname LT8\endcsname{\color{black}}%
    \fi
  \fi
  \setlength{\unitlength}{0.0500bp}%
  \begin{picture}(4438.00,2160.00)%
    \gplgaddtomacro\gplbacktext{%
      \csname LTb\endcsname%
      \put(791,15){\makebox(0,0)[r]{\strut{}\footnotesize $0$}}%
      \put(791,241){\makebox(0,0)[r]{\strut{}\footnotesize $0.0002$}}%
      \put(791,467){\makebox(0,0)[r]{\strut{}\footnotesize $0.0004$}}%
      \put(791,693){\makebox(0,0)[r]{\strut{}\footnotesize $0.0006$}}%
      \put(791,919){\makebox(0,0)[r]{\strut{}\footnotesize $0.0008$}}%
      \put(791,1145){\makebox(0,0)[r]{\strut{}\footnotesize $0.001$}}%
      \put(791,1371){\makebox(0,0)[r]{\strut{}\footnotesize $0.0012$}}%
      \put(791,1597){\makebox(0,0)[r]{\strut{}\footnotesize $0.0014$}}%
      \put(911,-185){\makebox(0,0){\strut{}}}%
      \put(1495,-185){\makebox(0,0){\strut{}}}%
      \put(2079,-185){\makebox(0,0){\strut{}}}%
      \put(2662,-185){\makebox(0,0){\strut{}}}%
      \put(3246,-185){\makebox(0,0){\strut{}}}%
      \put(3830,-185){\makebox(0,0){\strut{}}}%
      \put(4414,-185){\makebox(0,0){\strut{}}}%
      \put(91,890){\rotatebox{90}{\makebox(0,0){\strut{}\small Cell densities [pM]}}}%
      \put(2662,-285){\makebox(0,0){\strut{}}}%
      \put(2662,2066){\makebox(0,0){\strut{}\small Cell and factor profiles---extended model}}%
      \put(4326,1626){\makebox(0,0)[r]{\strut{}\footnotesize$t=100$ days}}%
    }%
    \gplgaddtomacro\gplfronttext{%
      \csname LTb\endcsname%
      \put(1764,1578){\makebox(0,0)[l]{\strut{}\footnotesize\oba}}%
      \csname LTb\endcsname%
      \put(1764,1378){\makebox(0,0)[l]{\strut{}\footnotesize\obp}}%
      \csname LTb\endcsname%
      \put(1764,1178){\makebox(0,0)[l]{\strut{}\footnotesize\oca}}%
      \csname LTb\endcsname%
      \put(1764,978){\makebox(0,0)[l]{\strut{}\footnotesize\ocm}}%
      \csname LTb\endcsname%
      \put(1764,778){\makebox(0,0)[l]{\strut{}\footnotesize\obu\ and \ocp}}%
    }%
    \gplbacktext
    \put(0,0){\includegraphics{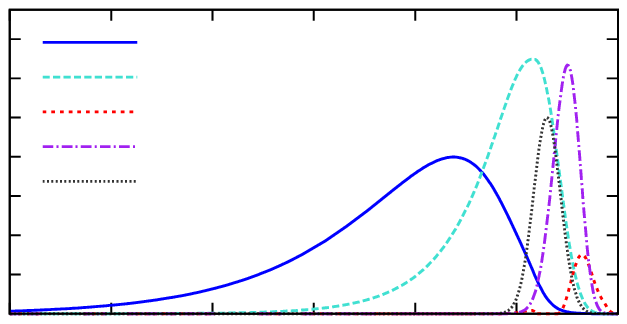}}%
    \gplfronttext
  \end{picture}%
\endgroup

%% file: profiles-model2-at-100d-factors.tex
\begingroup
  \makeatletter
  \providecommand\color[2][]{%
    \GenericError{(gnuplot) \space\space\space\@spaces}{%
      Package color not loaded in conjunction with
      terminal option `colourtext'%
    }{See the gnuplot documentation for explanation.%
    }{Either use 'blacktext' in gnuplot or load the package
      color.sty in LaTeX.}%
    \renewcommand\color[2][]{}%
  }%
  \providecommand\includegraphics[2][]{%
    \GenericError{(gnuplot) \space\space\space\@spaces}{%
      Package graphicx or graphics not loaded%
    }{See the gnuplot documentation for explanation.%
    }{The gnuplot epslatex terminal needs graphicx.sty or graphics.sty.}%
    \renewcommand\includegraphics[2][]{}%
  }%
  \providecommand\rotatebox[2]{#2}%
  \@ifundefined{ifGPcolor}{%
    \newif\ifGPcolor
    \GPcolortrue
  }{}%
  \@ifundefined{ifGPblacktext}{%
    \newif\ifGPblacktext
    \GPblacktexttrue
  }{}%
  \let\gplgaddtomacro\g@addto@macro
  \gdef\gplbacktext{}%
  \gdef\gplfronttext{}%
  \makeatother
  \ifGPblacktext
    \def\colorrgb#1{}%
    \def\colorgray#1{}%
  \else
    \ifGPcolor
      \def\colorrgb#1{\color[rgb]{#1}}%
      \def\colorgray#1{\color[gray]{#1}}%
      \expandafter\def\csname LTw\endcsname{\color{white}}%
      \expandafter\def\csname LTb\endcsname{\color{black}}%
      \expandafter\def\csname LTa\endcsname{\color{black}}%
      \expandafter\def\csname LT0\endcsname{\color[rgb]{1,0,0}}%
      \expandafter\def\csname LT1\endcsname{\color[rgb]{0,1,0}}%
      \expandafter\def\csname LT2\endcsname{\color[rgb]{0,0,1}}%
      \expandafter\def\csname LT3\endcsname{\color[rgb]{1,0,1}}%
      \expandafter\def\csname LT4\endcsname{\color[rgb]{0,1,1}}%
      \expandafter\def\csname LT5\endcsname{\color[rgb]{1,1,0}}%
      \expandafter\def\csname LT6\endcsname{\color[rgb]{0,0,0}}%
      \expandafter\def\csname LT7\endcsname{\color[rgb]{1,0.3,0}}%
      \expandafter\def\csname LT8\endcsname{\color[rgb]{0.5,0.5,0.5}}%
    \else
      \def\colorrgb#1{\color{black}}%
      \def\colorgray#1{\color[gray]{#1}}%
      \expandafter\def\csname LTw\endcsname{\color{white}}%
      \expandafter\def\csname LTb\endcsname{\color{black}}%
      \expandafter\def\csname LTa\endcsname{\color{black}}%
      \expandafter\def\csname LT0\endcsname{\color{black}}%
      \expandafter\def\csname LT1\endcsname{\color{black}}%
      \expandafter\def\csname LT2\endcsname{\color{black}}%
      \expandafter\def\csname LT3\endcsname{\color{black}}%
      \expandafter\def\csname LT4\endcsname{\color{black}}%
      \expandafter\def\csname LT5\endcsname{\color{black}}%
      \expandafter\def\csname LT6\endcsname{\color{black}}%
      \expandafter\def\csname LT7\endcsname{\color{black}}%
      \expandafter\def\csname LT8\endcsname{\color{black}}%
    \fi
  \fi
  \setlength{\unitlength}{0.0500bp}%
  \begin{picture}(4438.00,1828.00)%
    \gplgaddtomacro\gplbacktext{%
      \csname LTb\endcsname%
      \put(791,38){\makebox(0,0)[r]{\strut{}\footnotesize $0$}}%
      \put(791,264){\makebox(0,0)[r]{\strut{}\footnotesize $0.0002$}}%
      \put(791,490){\makebox(0,0)[r]{\strut{}\footnotesize $0.0004$}}%
      \put(791,716){\makebox(0,0)[r]{\strut{}\footnotesize $0.0006$}}%
      \put(791,942){\makebox(0,0)[r]{\strut{}\footnotesize $0.0008$}}%
      \put(791,1168){\makebox(0,0)[r]{\strut{}\footnotesize $0.001$}}%
      \put(791,1394){\makebox(0,0)[r]{\strut{}\footnotesize $0.0012$}}%
      \put(791,1620){\makebox(0,0)[r]{\strut{}\footnotesize $0.0014$}}%
      \put(911,-162){\makebox(0,0){\strut{}}}%
      \put(1495,-162){\makebox(0,0){\strut{}}}%
      \put(2079,-162){\makebox(0,0){\strut{}}}%
      \put(2662,-162){\makebox(0,0){\strut{}}}%
      \put(3246,-162){\makebox(0,0){\strut{}}}%
      \put(3830,-162){\makebox(0,0){\strut{}}}%
      \put(4414,-162){\makebox(0,0){\strut{}}}%
      \put(91,913){\rotatebox{90}{\makebox(0,0){\strut{}\small Concentrations (scaled)}}}%
      \put(2662,-262){\makebox(0,0){\strut{}}}%
    }%
    \gplgaddtomacro\gplfronttext{%
      \csname LTb\endcsname%
      \put(1764,1601){\makebox(0,0)[l]{\strut{}\footnotesize\opg}}%
      \csname LTb\endcsname%
      \put(1764,1401){\makebox(0,0)[l]{\strut{}\footnotesize\rankl}}%
      \csname LTb\endcsname%
      \put(1764,1201){\makebox(0,0)[l]{\strut{}\footnotesize\tgfb}}%
    }%
    \gplbacktext
    \put(0,0){\includegraphics{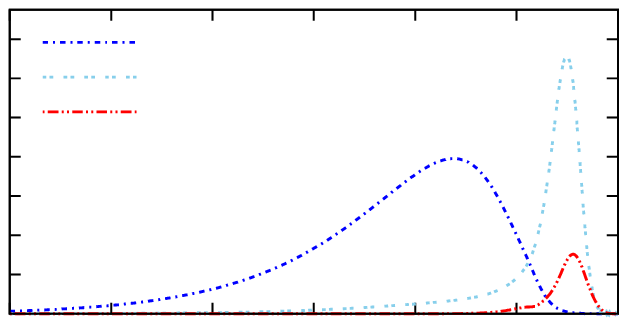}}%
    \gplfronttext
  \end{picture}%
\endgroup

%% file: bmu-cavity-sketch-to-3000.tex
\begingroup
  \makeatletter
  \providecommand\color[2][]{%
    \GenericError{(gnuplot) \space\space\space\@spaces}{%
      Package color not loaded in conjunction with
      terminal option `colourtext'%
    }{See the gnuplot documentation for explanation.%
    }{Either use 'blacktext' in gnuplot or load the package
      color.sty in LaTeX.}%
    \renewcommand\color[2][]{}%
  }%
  \providecommand\includegraphics[2][]{%
    \GenericError{(gnuplot) \space\space\space\@spaces}{%
      Package graphicx or graphics not loaded%
    }{See the gnuplot documentation for explanation.%
    }{The gnuplot epslatex terminal needs graphicx.sty or graphics.sty.}%
    \renewcommand\includegraphics[2][]{}%
  }%
  \providecommand\rotatebox[2]{#2}%
  \@ifundefined{ifGPcolor}{%
    \newif\ifGPcolor
    \GPcolortrue
  }{}%
  \@ifundefined{ifGPblacktext}{%
    \newif\ifGPblacktext
    \GPblacktexttrue
  }{}%
  \let\gplgaddtomacro\g@addto@macro
  \gdef\gplbacktext{}%
  \gdef\gplfronttext{}%
  \makeatother
  \ifGPblacktext
    \def\colorrgb#1{}%
    \def\colorgray#1{}%
  \else
    \ifGPcolor
      \def\colorrgb#1{\color[rgb]{#1}}%
      \def\colorgray#1{\color[gray]{#1}}%
      \expandafter\def\csname LTw\endcsname{\color{white}}%
      \expandafter\def\csname LTb\endcsname{\color{black}}%
      \expandafter\def\csname LTa\endcsname{\color{black}}%
      \expandafter\def\csname LT0\endcsname{\color[rgb]{1,0,0}}%
      \expandafter\def\csname LT1\endcsname{\color[rgb]{0,1,0}}%
      \expandafter\def\csname LT2\endcsname{\color[rgb]{0,0,1}}%
      \expandafter\def\csname LT3\endcsname{\color[rgb]{1,0,1}}%
      \expandafter\def\csname LT4\endcsname{\color[rgb]{0,1,1}}%
      \expandafter\def\csname LT5\endcsname{\color[rgb]{1,1,0}}%
      \expandafter\def\csname LT6\endcsname{\color[rgb]{0,0,0}}%
      \expandafter\def\csname LT7\endcsname{\color[rgb]{1,0.3,0}}%
      \expandafter\def\csname LT8\endcsname{\color[rgb]{0.5,0.5,0.5}}%
    \else
      \def\colorrgb#1{\color{black}}%
      \def\colorgray#1{\color[gray]{#1}}%
      \expandafter\def\csname LTw\endcsname{\color{white}}%
      \expandafter\def\csname LTb\endcsname{\color{black}}%
      \expandafter\def\csname LTa\endcsname{\color{black}}%
      \expandafter\def\csname LT0\endcsname{\color{black}}%
      \expandafter\def\csname LT1\endcsname{\color{black}}%
      \expandafter\def\csname LT2\endcsname{\color{black}}%
      \expandafter\def\csname LT3\endcsname{\color{black}}%
      \expandafter\def\csname LT4\endcsname{\color{black}}%
      \expandafter\def\csname LT5\endcsname{\color{black}}%
      \expandafter\def\csname LT6\endcsname{\color{black}}%
      \expandafter\def\csname LT7\endcsname{\color{black}}%
      \expandafter\def\csname LT8\endcsname{\color{black}}%
    \fi
  \fi
  \setlength{\unitlength}{0.0500bp}%
  \begin{picture}(4438.00,886.00)%
    \gplgaddtomacro\gplbacktext{%
      \csname LTb\endcsname%
      \put(2662,27){\makebox(0,0){\strut{}\small Distance from \bmu\ front [$\um$]}}%
    }%
    \gplgaddtomacro\gplfronttext{%
      \csname LTb\endcsname%
      \put(791,585){\makebox(0,0)[r]{\strut{}\footnotesize $-100$}}%
      \csname LTb\endcsname%
      \put(791,702){\makebox(0,0)[r]{\strut{}\footnotesize $0$}}%
      \csname LTb\endcsname%
      \put(791,819){\makebox(0,0)[r]{\strut{}\footnotesize $100$}}%
      \csname LTb\endcsname%
      \put(911,327){\makebox(0,0){\strut{}\footnotesize $-3000$}}%
      \csname LTb\endcsname%
      \put(1495,327){\makebox(0,0){\strut{}\footnotesize $-2500$}}%
      \csname LTb\endcsname%
      \put(2079,327){\makebox(0,0){\strut{}\footnotesize $-2000$}}%
      \csname LTb\endcsname%
      \put(2662,327){\makebox(0,0){\strut{}\footnotesize $-1500$}}%
      \csname LTb\endcsname%
      \put(3246,327){\makebox(0,0){\strut{}\footnotesize $-1000$}}%
      \csname LTb\endcsname%
      \put(3830,327){\makebox(0,0){\strut{}\footnotesize $-500$}}%
      \csname LTb\endcsname%
      \put(4414,327){\makebox(0,0){\strut{}\footnotesize $0$}}%
    }%
    \gplbacktext
    \put(0,0){\includegraphics{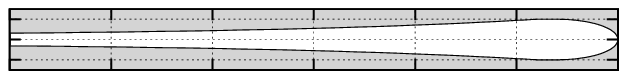}}%
    \gplfronttext
  \end{picture}%
\endgroup

%% file: moving-bmu.bbl
\begin{thebibliography}{99}
    \setlength{\itemsep}{-0.3ex}\small
    \bibitem{martin-burr-sharkey} Martin R B, Burr D B and Sharkey N A 1998 \textit{Skeletal Tissue Mechanics} (New York: Springer)
    \bibitem{bone-mechanics-handbook} ed S C Cowin 2001 \textit{Bone Mechanics Handbook} $2^    \text{nd}$Ed (Boca Raton: CRC Press)
    \bibitem{frost2} Frost H M 1983 The skeletal intermediary organization. \textit{Metab. Bone Dis. \& Rel. Res.} \textbf{4}:281--290
    \bibitem{frost3} Frost H M 1964 Dynamics of bone remodeling. In \textit{Bone Biodynamics}, ed H M Frost (Boston: Little, Brown \& Co.)
    \bibitem{parfitt1} Parfitt A M 1979 Quantum concept of bone remodeling and turnover: implications for the pathogenesis of osteoporosis. \emph{Calcif. Tissue Int.} \textbf{28}:1--5
    \bibitem{parfitt2} Parfitt A M 1984 The cellular basis of bone remodeling: the quantum concept reexamined in light of recent advances in the cell biology of bone. \emph{Calcif. Tissue Int.} \textbf{36}:S37--S45
    \bibitem{parfitt3} Parfitt A M 1994 Osteonal and hemi-osteonal remodeling: the spatial and temporal framework for signal traffic in adult human bone. \textit{J. Cell. Biochem.} \textbf{55}:273--286
	\bibitem{jaworski-hooper} Jaworski Z F G and Hooper C 1980. Study of cell kinetics within evolving secondary haversian systems. \emph{J.~Anat.~London} {\bf 131}:91--102
    \bibitem{jaworski-duck-sekaly} Jaworski Z F G, Duck B and Sekaly G 1981. Kinetics of osteoclasts and their nuclei in evolving secondary Haversian systems. \textit{J. Anat.} \textbf{133}:397--405
    \bibitem{vaananen-etal} V\"a\"an\"anen H K, Liu Y-K, Lehenkari P and Uemara T 1998 How do osteoclasts resorb bone? \textit{Mater. Sci. and Eng. C} \textbf{6}:205--209
    \bibitem{roodman} Roodman G D 1999 Cell biology of the osteoclast. \textit{Exp. Hematology} \textbf{27}:1229--1241
    \bibitem{martin} Martin T J 2004 Paracrine regulation of osteoclast formation and activity: Milestones in discovery. \textit{J. Musculoskel. Neuron. Interact.} \textbf{4}:243--253
    \bibitem{lemaire-etal} Lemaire V, Tobin F L, Greller L D, Cho C R and Suva L J 2004 Modeling the interactions between osteoblast and osteoclast activities in bone remodeling \textit{J. Theor. Biol.} \textbf{29}:293--309
	\bibitem{pivonka-etal1} Pivonka P, Zimak J, Smith D W, Gardiner B S, Dunstan C R, Sims N A, Martin T J and Mundy G R 2008 Model structure and control of bone remodeling:         a theoretical study. Bone {\bf 43}:249--263
    \bibitem{komarova-etal1} Komarova S V, Smith R J, Dixon S J, Sims S M and Wahl L M 2003 Mathematical model predicts a critical role for osteoclast autocrine regulation in the control of bone remodeling \textit{J. Theor. Biol.} \textbf{229}:293--309
    \bibitem{ryser-etal1} Ryser M D, Nigam N and Komarova S V 2009 Mathematical modeling of spatio-temporal dynamics of a single bone multicellular unit. \emph{J. Bone Miner. Res.} \textbf{24}:860--870
    \bibitem{ryser-etal2} Ryser M D, Komarova S V and Nigam N 2010. The cellular dynamics of bone remodelling: a mathematical model. \textit{SIAM J. Appl. Math.} \textbf{70}:1899--1921
    \bibitem{huiskes-etal1} van Oers R F M, Ruimerman R, Tanck E, Hilbers P A J and Huiskes R 2008 A unified theory for osteonal and hemi-osteonal remodeling \textit{Bone} \textbf{42}:250--259
    \bibitem{stewart-lightfoot} Bird R B, Stewart W E and Lightfoot E N 2002 \emph{Transport Phenomena} $2^\text{nd}$Ed (New York: Wiley)
    \bibitem{evans-morriss} Evans D J and Morriss G 2008 \textit{Statistical mechanics of nonequilibrium liquids} $2^\text{nd}$Ed (Cambridge University Press)
    \bibitem{scheurer-stueckelberg} Scheurer P B and Stueckelberg de Breidenbach E C G 1974 \emph{Thermocin\'etique Ph\'enom\'enologique Galil\'eenne} (Basel: Birkh\"auser)
    \bibitem{bronckers-etal} Bronckers A L J J, Goel W, Luo G, Karsenty G, D'Souza R N, Lyaruu D M and Burger E H 1996 DNA fragmentation during bone formation in neonatal rodents assessed by transferase-mediated end labeling. \textit{J. Bone Miner. Res.} \textbf{11}:1281--1291
    \bibitem{lauffenburger-linderman} Lauffenburger D A and Linderman J J 1993 \textit{Receptors: models for binding, trafficking, and signaling.} (New York: Oxford Univ. Press)    
    \bibitem{burger-etal} Burger E H, Klein-Nulend J and Smit T H 2003 Strain-derived canalicular fluid flow regulates osteoclast activity in a remodelling osteon---a proposal. \textit{J. Biomech.} \textbf{36}:1453--1459
    \bibitem{ishii-etal1} Ishii M, Egen J G, Klauschen F, Meier-Schellersheim M, Saeki Y, Vacher J, Proia R L and Germain R N 2009 Sphingosine-1-phosphate mobilizes osteoclast precursors and regulates bone homeostasis. \textit{Nature} \textbf{458}:524--529
    \bibitem{ishii-etal2} Ishii T, Kikuta J, Kubo A and Ishii M 2010 Control of osteoclast precursor migration: A novel point of control for osteoclastogenesis and bone homeostasis. \textit{IBMS BoneKEy} \textbf{7} 279--286
    \bibitem{harada-rodan} Harada S-I and Rodan G A 2003 Control of osteoblast function and regulation of bone mass. \textit{Nature} \textbf{423}:349--355
    \bibitem{iqbal-sun-zaidi} Iqbal J, Sun L and Zaidi M 2009 Coupling bone degradation to formation. \textit{Nat. Med.} \textbf{15}:729--731
    \bibitem{tang-etal} Tang Y \etal\ 2009 \Tgfb 1-induced migration of bone mesenchymal stem cells couples bone resorption with formation. \textit{Nat. Med.} \textbf{15}:757--766
    \bibitem{gori-etal} Gori F, Hofbauer L C, Dunstan C R, Spelsberg T C, Kholsa S and Riggs B L 2000 The expression of osteoprotegerin and \rank\ ligand and the support of osteoclast formation by stromal-osteoblast linearge cells is developmentally regulated. \textit{Endocrinology} \textbf{141}:4768--4776
    \bibitem{thomas-etal} Thomas G P, Baker S U, Eisman J A and Gardiner E M 2001 Changing \rankl/\opg\ mRNA expression in differentiating murine primary osteoblasts. \textit{J. Endocrinol.} \textbf{170}:451--460
\bibitem{mathematica} Wolfram Research, Inc. 2008 \textit{Mathematica Edition: Version 7.0} (Champaign, Illinois: Wolfram Research, Inc.)
\bibitem{buenzli-pivonka-etal} Buenzli P R, Pivonka P, Gardiner B S, Smith D W, Dunstan C R and Mundy G R 2010 Theoretical analysis of the spatio-temporal structure of bone multicellular units. Proceedings of the WCCM/APCOM 2010, Sydney, \textit{IOP Conf. Ser.: Mater. Sci. Eng.} \textbf{10}:012132
    \bibitem{halmos} Halmos P R 1974 \textit{Measure theory} (New York: Springer)
\end{thebibliography}
